\documentclass[12pt, a4paper]{article}

\usepackage{a4,bm}
\usepackage{parskip}
\usepackage{hyperref}
\usepackage{color,soul,rotating}
\usepackage{xcolor}

\usepackage{bm}
\usepackage{braket}
\usepackage{amssymb,amsmath,amsfonts,amsxtra,amscd}
\usepackage{amsthm}
\theoremstyle{remark}      
\newtheorem{rem}{Remark}   
\newtheorem{hyp}{Hypothesis}
\newtheorem{thm}{Theorem}
\newtheorem{lem}{Lemma}
\newtheorem{cor}{Corollary}
\theoremstyle{definition}
\newtheorem{defn}{Definition}[section]
\newtheorem{prop}{Proposition}
\newtheorem{conj}{Conjecture}

\usepackage{physics}
\usepackage{graphicx}
\usepackage{float}
\usepackage{wrapfig}
\usepackage{mwe}
\usepackage{cancel}
\usepackage{longtable}
\usepackage{tabularx}

\usepackage[a4paper]{geometry}
\usepackage{empheq}
\usepackage[theorems,skins]{tcolorbox}

\usepackage{tikz}
\usepackage{tikz-cd}
\usepackage{lmodern}
\usepackage[framemethod=TikZ]{mdframed}
\usepackage{youngtab}
\usepackage{ytableau}
\usepackage{tikzsymbols}
\usepackage{textcomp}
\usetikzlibrary{calc} 
\usetikzlibrary{matrix,arrows,decorations.pathmorphing}
\usetikzlibrary{positioning,bending}
\usetikzlibrary{decorations.text}
\usetikzlibrary{decorations.markings}
\usetikzlibrary{snakes}
\tikzset{>=stealth}
\tikzset{snake it/.style={decorate, decoration=snake}}

\usepackage[skins,theorems]{tcolorbox}
\tcbset{highlight math style={enhanced,colframe=red,colback=green! 4!,arc=2pt,boxrule=1pt}}

\usepackage{xcolor}
\definecolor{bostonuniversityred}{rgb}{0.8, 0.0, 0.0}
\definecolor{brightmaroon}{rgb}{0.76, 0.13, 0.28}
\usepackage{geometry}
\usepackage{mathtools}
\usepackage{graphicx}
\usepackage{enumitem}

\numberwithin{equation}{section}

\title{Cohomological Aspects of Entanglement Entropy:
	From Information Theory to Noncommutative Geometry}

\author{R.~C.~Rashkov\footnote{email:rash@phys.uni-sofia.bg;\, rash@hep.itp.tuwien.ac.at} \\
	\\
	\textit{Department of Physics, Sofia University,}\\    {5 J. Bourchier Blvd., 1164 Sofia, Bulgaria} \\
	and\\
	\textit{ITP, Vienna University of Technology,}\\
	{Wiedner Hauptstr. 8--10, 1040 Vienna, Austria}
}

\date{}

\begin{document}
	
	\maketitle	
\begin{abstract}
	We develop a cohomological framework for entanglement entropy that
	unifies information theory, operator algebras, and noncommutative
	geometry. Starting from the information-theoretic characterization
	of entropy as a $1$-cocycle, we pass to the quantum setting via
	Hochschild and cyclic cohomology. For an inclusion of von Neumann
	algebras $\mathcal{N}\subset\mathcal{M}$ with a faithful normal
	state $\omega$ admitting a cyclic and separating vector, and a
	conditional expectation $E:\mathcal{M}\to\mathcal{N}$ whose
	existence is equivalent to modular invariance of $\mathcal{N}$
	(Takesaki's theorem), we construct the entanglement complex
	$C^\bullet_{\mathrm{ent}}(\mathcal{A},\mathcal{B},\omega)$ as the
	image of the transgression map $\tau_\omega$ on the relative
	cyclic complex $C^\bullet_\lambda(\mathcal{A},\mathcal{B})$. Our
	main result is that, under explicit analytic hypotheses on the
	relative modular cocycle, the inclusion
	$C^\bullet_{\mathrm{ent}}\hookrightarrow
	C^\bullet_\lambda(\mathcal{A},\mathcal{B})$ is a quasi-isomorphism,
	which identifies the entanglement cohomology with the kernel of
	the restriction map $HC^\bullet(\mathcal{A})\to
	HC^\bullet(\mathcal{B})$, up to the connecting homomorphism
	$\delta$ of the long exact sequence of the pair. The relative
	modular cocycle generates explicit nonzero entanglement classes
	whenever $E$ is not $\omega$-invariant, and the Araki relative
	entropy is its infinitesimal generator. The framework accommodates
	the Type III von Neumann algebras of quantum field theory, where no
	local density matrix exists, and it leads to a holographic
	proposal: a distinguished subspace of $H^1_{\mathrm{ent}}$ is
	conjecturally dual to the space of linearized modular Hamiltonian
	perturbations of the bulk entanglement wedge. For a spherical
	entangling region perturbed by a primary scalar operator, we
	propose a conjectural general-$d$ formula for the coefficient
	$\kappa_d(\Delta)$ of the smeared-source relative entropy; the
	companion Letter verifies the corresponding JLMS-type equality for
	point-like sources in $d=2,3,4$, a related but distinct
	computation. Worked examples in finite dimensions, type II$_1$
	factors, and type III$_1$ factors illustrate the construction.
\end{abstract}

	\tableofcontents
	
\section{Introduction}
\label{sec:intro}

\subsection{Overview and Motivation}
\label{subsec:intro-1}

Entanglement has become one of the most pervasive concepts in modern
physics, extending far beyond its origins in quantum information theory.
It now plays central roles in condensed matter physics (where it
characterizes quantum phases and topological order), in general
relativity (where it underpins the holographic principle), and in
high-energy physics (where it provides insights into the structure of
quantum field theory and the black hole information paradox).

This remarkable ubiquity suggests that entanglement is not merely a
computational tool but a fundamental structural feature of quantum
theory. Indeed, the sophisticated physics of entanglement frequently
requires equally sophisticated mathematics, which in turn reveals
unexpected features and deep connections between apparently disparate
areas.

The central observation that motivates this work is the structural
similarity between the chain rule of Shannon entropy and the Leibniz
rule for derivatives. In the framework of ``Information Cohomology''
developed by Baudot and Bennequin \cite{baudot}, this analogy is made
precise: entropy is a \(1\)-cocycle in a cohomology theory on the
category of probability spaces. Mutual information, the failure of
entropy to be additive, is a coboundary. Higher-order information
measures correspond to higher-degree cohomology classes.  The most striking result of the Baudot-Bennequin framework is a classification theorem: When you compute the first cohomology group $H^1$ of
an information structure with appropriate probabilistic coefficients, the
solutions to the 1-cocycle equations are not arbitrary. Under mild non-degeneracy and continuity conditions, every 1-cocycle is locally a real multiple of the Shannon entropy (or its parameterized generalizations, such
as Tsallis and Rényi entropies, depending on the choice of module coefficients $\mathcal{F}_\alpha$).In other words, you do not need to impose entropy by hand like as Faddeev did\cite{D-Faddeev}; entropy appears spontaneously as the unique non-trivial generator (cohomology class) of the information complex.

Translating this insight to the quantum realm, where von Neumann
entropy replaces Shannon entropy, requires a framework that
accommodates the noncommutative structure of quantum observables. This
is precisely what Hochschild and cyclic cohomology provide. The key
idea is that entanglement entropy, as a measure of quantum
correlations, should be understood as a cohomological obstruction: the
failure of the restriction map from the full algebra of observables to
a subalgebra representing a subsystem.

In this work, we develop this cohomological framework for entanglement
entropy, unifying perspectives from information theory, operator
algebras, and noncommutative geometry. Our main results are:
\begin{itemize}
	\item A precise identification of von Neumann entropy as a
	\(1\)-cocycle in the relative cyclic cohomology of an inclusion
	of von Neumann algebras \(\mathcal{N} \subset \mathcal{M}\), with
	the modular structure governed by Takesaki's theorem on the
	existence of the conditional expectation \(E: \mathcal{M} \to
	\mathcal{N}\).
	
		\item The construction of an entanglement complex
	\(C^\bullet_{\mathrm{ent}}(\mathcal{A}, \mathcal{B}, \omega)\),
	defined as the image of the transgression map \(\tau_\omega\) on
	the relative cyclic complex
	\(C^\bullet_\lambda(\mathcal{A}, \mathcal{B})\). Its cohomology
	\(H^\bullet_{\mathrm{ent}}(\mathcal{A}, \mathcal{B}, \omega)\)
	classifies entanglement structures, and its first cohomology
	contains the classes generated by the relative modular cocycle
	\(u^E\).
	
\item The proof that, under the analytic Hypotheses
	\ref{hyp:cocycle-in-A} and~\ref{hyp:growth} of
	Section~\ref{subsec:prelim-5}, the inclusion
	\(C^\bullet_{\mathrm{ent}} \hookrightarrow
	C^\bullet_\lambda(\mathcal{A}, \mathcal{B})\) is a
	quasi-isomorphism, which identifies entanglement cohomology with
	the kernel of the restriction map
	\(HC^\bullet(\mathcal{A}) \to HC^\bullet(\mathcal{B})\), up to the
	connecting homomorphism \(\delta\) of the long exact sequence of
	the pair \((\mathcal{A}, \mathcal{B})\).
	
	\item A demonstration that the Connes--Radon--Nikodym cocycle
	\(u^E_t = (D\omega : D(\omega|_{\mathcal{N}}) \circ E)_t\), the
	fundamental object of Tomita--Takesaki modular theory, encodes the
	dynamical structure of relative entanglement, generalizing the
	density matrix to Type III algebras where no trace exists.
	
		\item A holographic proposal (Conjecture
	\ref{conj:holographic} in Section \ref{subsubsec:ent-holographic})
	identifying a distinguished subspace of
	\(H^1_{\mathrm{ent}}\) with the space of linearized modular
	Hamiltonian perturbations of the bulk entanglement wedge, with the
	Araki relative entropy as the quantitative bridge. The proposal
	includes a conjectural general-\(d\) formula for the coefficient
	$\kappa_d(\Delta)$ of the smeared-source relative entropy; the
	companion Letter~\cite{Rashkov:2026letter} verifies the JLMS-type
	equality for point-like sources in $d=2,3,4$.
\end{itemize}

This framework naturally accommodates the Type III von Neumann algebras
that appear in quantum field theory, where no local density matrix
exists and entanglement cannot be defined by tracing out tensor
factors.

\subsection{Holographic Entanglement Entropy}
\label{subsec:intro-2}

\subsubsection*{The Ryu--Takayanagi Formula}

One of the most remarkable incarnations of entanglement in recent
physics is holographic entanglement entropy (HEE). The Ryu--Takayanagi
formula \cite{Ryu:2006bv} provides a geometric dual to entanglement
entropy in the context of the AdS/CFT correspondence. For a static
state in a CFT\(_d\) dual to classical gravity in AdS\(_{d+1}\), the
entanglement entropy \(S_A\) of a boundary spatial region \(A\) is
given by:
\begin{equation}
	S_A = \frac{\text{Area}(\gamma_A)}{4G_N}
	\label{eq:RT}
\end{equation}
where \(\gamma_A\) is the minimal-area extremal surface in the bulk AdS
that is homologous to \(A\) (i.e., \(\partial \gamma_A = \partial A\)),
and \(G_N\) is Newton's constant in the bulk.

This formula can be viewed as a generalization of the Bekenstein--Hawking
black hole entropy to general quantum states, identifying entanglement
entropy with a geometric measure in the gravitational dual.

\subsubsection*{Extensions and Generalizations}

The Ryu--Takayanagi formula has been extended in several important
directions:

- HRT Formula: Hubeny, Rangamani, and Takayanagi \cite{Hubeny:2007xt}
generalized the formula to dynamical spacetimes by replacing the
minimal surface with an extremal surface anchored to the boundary at
the same time slice as \(A\).

- Quantum Corrections: Subleading corrections take the form
\(S_A = \frac{\text{Area}}{4G_N} + S_{\text{bulk}}\), where
\(S_{\text{bulk}}\) is the entanglement entropy of bulk quantum fields
across \(\gamma_A\).

- Islands: In evaporating black hole contexts, the entanglement entropy
of radiation includes contributions from ``islands''---regions inside
the black hole connected via quantum extremal surfaces---resolving the
black hole information paradox.

\textit{Physical Implications}: The physical implications of these
findings have significantly deepened our understanding of fundamental
physics:

- Emergent Spacetime: HEE suggests that spacetime connectivity emerges
from entanglement structure (the ``ER = EPR'' conjecture).

- Entanglement Wedge Reconstruction: The bulk region bounded by
\(A \cup \gamma_A\) (the entanglement wedge) is dual to the boundary
region \(A\), establishing holographic subregion duality.

- Quantum Information in Gravity: The quantum extremal surface formula
incorporates bulk entropy, capturing fine-grained quantum information.

Holographic entanglement entropy provides a powerful geometric
dictionary for quantum information in gravity, linking entanglement,
geometry, and emergent spacetime in a unified framework.

\textit{Remark}: For comprehensive reviews of holographic entanglement
entropy and its applications, see \cite{Nishioka:2018khk} and
\cite{Witten:2018zxz}. For recent developments connecting von Neumann
algebras to emergent spacetime, see \cite{Liu:2025krl}.

\subsubsection*{Relation to the Present Work}

The present work does not attempt to reproduce the Ryu--Takayanagi
formula from cohomology. Rather, it provides a finer invariant---the
entanglement cohomology of Section \ref{sec:ent-cohom}---that detects
the modular structure of the state, not merely its entropy. The
holographic interpretation of this finer invariant is discussed in
Section \ref{subsubsec:ent-holographic}. In brief, the first
cohomology \(H^1_{\mathrm{ent}}\) is conjecturally dual to the space of
linearized modular Hamiltonian perturbations of the entanglement
wedge, with the Araki relative entropy providing the quantitative
bridge between the boundary CRN cocycle and the bulk relative entropy
between geometries.

\subsection{Entanglement in Quantum Field Theory}
\label{subsec:intro-3}

\subsubsection*{The Challenge of Type III Algebras}

In quantum field theory, local observable algebras
\(\mathcal{A}(\mathcal{O})\) are typically Type III factors
\cite{Araki-64}. This has profound consequences for entanglement:
\begin{itemize}
	\item No Trace: There is no trace on a Type III factor, so there is
	no local density matrix \(\rho_\mathcal{O}\).
	
	\item Divergent Entropy: The von Neumann entropy
	\(S(\rho_\mathcal{O})\) diverges for any state.
	
	\item No Tensor Factorization: The algebra of a region and its
	complement do not form a tensor product.
\end{itemize}

As a result, entanglement cannot be defined in QFT by the standard
method of tracing out tensor factors. This is a major obstacle that
must be overcome.

\subsubsection*{The Araki Resolution: Modular Theory}

The resolution, pioneered by Araki \cite{Araki-64} and developed by
Connes \cite{Connes-73}, is to replace the density matrix with the
modular theory of Tomita and Takesaki. For a faithful normal state
\(\omega\) on a von Neumann algebra \(\mathcal{M}\), the modular
operator \(\Delta_\omega\) and modular automorphism group
\(\sigma_t^\omega(x) = \Delta_\omega^{it} x \Delta_\omega^{-it}\)
encode the intrinsic ``thermal'' and entanglement structure of the
state.

Key facts:

- The relative entropy \(S(\phi \| \omega)\) is defined even though
individual entropies may diverge.

- The Connes--Radon--Nikodym cocycle \((D\phi : D\omega)_t\) provides a
rigorous replacement for the density matrix ratio.

- The modular Hamiltonian \(K = -\log \Delta_\omega\) is the
``entanglement Hamiltonian'' even when no trace exists.

\subsubsection*{The Role of Hochschild and Cyclic Cohomology}

The Tomita--Takesaki modular theory provides the data (modular
operator, modular automorphism group, CRN cocycle) that we need to
construct a cohomological theory of entanglement. The key insight is
that:

- The modular Hamiltonian defines a derivation
\(\delta(a) = [K, a]\), which is a Hochschild \(1\)-cocycle.

- The CRN cocycle satisfies a chain rule, making it a cyclic cocycle.

- The relative entropy is the value of this cocycle on the state.

This is the quantum analog of the Information Cohomology framework of
Baudot and Bennequin \cite{baudot}, adapted to the noncommutative
setting of von Neumann algebras.

\subsection{Information Theory and Categorical Structures}
\label{subsec:intro-4}

\subsubsection*{Why Categories?}

When dealing with complicated quantum systems, an important step is to
organize the available information in a useful way. Categorical
approaches provide a general framework for this organization, allowing
us to:
\begin{itemize}
	\item Distinguish between different types of information structures.
	\item Define operations that combine and transform information.
	\item Identify universal properties and invariants.
	\item Connect to cohomology via functorial constructions.
\end{itemize}

The categorical perspective is particularly powerful because it reveals
that many information-theoretic quantities (entropy, mutual
information, conditional entropy) are actually cohomological invariants
of the underlying category of observables.

\subsubsection*{Information Categories}

Following Vigneaux \cite{vigneaux}, we define an information category
\(\mathcal{C}\) as a set of random variables defined on a common
probability space, satisfying:

a) Closure under coarsening: If \(X \in \mathcal{C}\) and \(Y = f(X)\)
for some function \(f\), then \(Y \in \mathcal{C}\).

b) Closure under joins: If \(X, Y \in \mathcal{C}\), then their join
\(X \vee Y\) (the random variable corresponding to the coarsest common
refinement of their partitions) is in \(\mathcal{C}\).

c) Contains the trivial variable: The trivial random variable
\(\mathbf{1}\) (constant everywhere) belongs to \(\mathcal{C}\).

These conditions ensure that \(\mathcal{C}\) is closed under the
operations that generate new information from existing information.

\subsubsection*{The Functorial Structure}

To make this framework operational, we consider a poset \(S\)
consisting of objects/observables and a functor \(E\) assigning to
observables their possible values \cite{vigneaux}. This organizes all
information quantities into a pair of categories:

- Category I: Generates an algebraic structure like a monoid, or more
generally a monad. This encodes the operations on information (e.g.,
combining variables, conditioning).

- Category II: Provides a representation of the algebraic structure.
This encodes the actual values and distributions.

Physical Interpretation: Category I represents the ``syntax'' of
information (how to combine and manipulate information), while Category
II represents the ``semantics'' (the actual information content).

\subsubsection*{The Measure Problem}

A central problem in any quantum theory (and not only) is the measure
problem: how to quantify the amount of information or uncertainty in a
system. In the categorical context, this is treated as follows.

Let \(S = \{e_1, e_2, \ldots, e_n\}\) be a finite choice system and
\(p_i = p(e_i)\) a probability distribution over it. The measure of
uncertainty \(H(p_1, p_2, \ldots, p_n)\) is entirely defined by the
following conditions \cite{baudot}:

H1 (Symmetry):
\(H(p_1, \ldots, p_n) = H(p_{\sigma(1)}, \ldots, p_{\sigma(n)})\) for
any permutation \(\sigma\). The measure does not depend on the
numbering of the possible choices.

H2 (Continuity): \(H(p_1, \ldots, p_n)\) is a continuous function in
all variables. Small changes in probabilities cause small changes in
entropy.

H3 (Grouping Axiom): The chaining formula
\begin{equation}
	H(p_1, p_2, \ldots, p_n) = H(p_1 + p_2, p_3, \ldots, p_n) +
	(p_1 + p_2) H\left(\frac{p_1}{p_1+p_2}, \frac{p_2}{p_1+p_2}\right)
\end{equation}
This states that the entropy of a partition can be computed by first
grouping some outcomes and then computing the entropy within the group.

H4 (Monotonicity): For a uniform probability distribution
\(p_i = 1/n\), the entropy \(H(1/n, \ldots, 1/n)\) as a function of
\(n\) is monotone increasing. With uniform distribution, the
uncertainty should increase with the number of possibilities.

\subsubsection*{The Fundamental Theorems}

\textit{Theorem A}: (H1)--(H4) are satisfied if, and only if,
\begin{equation}
	H(p_1, p_2, \ldots, p_n) = -\sum_{i=1}^n p_i \log p_i
\end{equation}
This is the Shannon entropy \cite{Shannon:1948}.

\textit{Theorem B}: Let \(\alpha > 0\) and
\(u: [0,1] \to \mathbb{R}\) be a measurable function which satisfies
the Fundamental Equation of Information Theory:
\begin{align}
	&\forall (x,y) \in [0,1)^2 \quad \text{such that } x + y \in [0,1]
	\nonumber \\
	& u(1-x) + (1-x)^\alpha u\left(\frac{y}{1-x}\right) =
	u(y) + (1-y)^\alpha u\left(\frac{1-x-y}{1-y}\right)
\end{align}

Then there exists a real number \(\lambda\) such that for all
\(x \in [0,1]\):
\begin{equation}
	u(x) = \begin{cases}
		\lambda \left(-x\log_2(x) - (1-x)\log_2(1-x)\right)
		= \lambda S_1(x,1-x), & \text{if } \alpha = 1 \\[12pt]
		\lambda \left(\frac{1}{1-\alpha}(x^\alpha + (1-x)^\alpha - 1)\right)
		= \lambda S_\alpha(x,1-x), & \text{if } \alpha \neq 1
	\end{cases}
\end{equation}

This theorem shows that the Shannon entropy (and its generalization,
the Tsallis entropy) are essentially unique measures of uncertainty
satisfying the fundamental equation.

\subsubsection*{From Classical to Quantum Entropy}

These theorems have a direct quantum analog. In quantum mechanics, the
von Neumann entropy:
\begin{equation}
	S(\rho) = -\operatorname{Tr}(\rho \log \rho)
\end{equation}
satisfies analogous axioms, with the Shannon entropy replaced by the
von Neumann entropy and the probability distribution replaced by the
eigenvalues of the density matrix.

The connection to cohomology is as follows:

- In the Information Cohomology framework of Baudot and Bennequin
\cite{baudot}, the Shannon entropy is a \(1\)-cocycle.

- The quantum analog, von Neumann entropy, is a \(1\)-cocycle in the
Hochschild cohomology of the algebra of observables.

- Mutual information (classical) and mutual information (quantum) are
coboundaries---they measure the failure of additivity.

- The fundamental equation of information theory is the cocycle
condition in this cohomology.

This will be the starting point for our cohomological framework in
Sections \ref{sec:cohom} and \ref{sec:ent-cohom}.

The key insight is that entropy is not just a statistical quantity but
a topological invariant---a \(1\)-cocycle that measures the obstruction
to decomposing a system into independent parts. This is precisely what
entanglement entropy measures in the quantum setting.

\subsection{Overview of the Paper}
\label{subsec:intro-5}

\subsubsection*{Organization}

The paper is organized as follows.

Section \ref{sec:prelim} reviews the standard formalism of multipartite
quantum systems, including Hilbert spaces, density matrices, and the
geometric characterization of entangled states via the Segre embedding.
It then introduces the operator-algebraic framework that will be used
throughout the rest of the paper: the von Neumann algebra pair
\(\mathcal{N} \subset \mathcal{M}\), the faithful normal state
\(\omega\) with cyclic and separating vector \(\Omega_\omega\), the
conditional expectation \(E: \mathcal{M} \to \mathcal{N}\) whose
existence is equivalent to modular invariance of \(\mathcal{N}\)
(Takesaki's theorem), the relative modular cocycle \(u^E\), and the
dense \(*\)-subalgebra \(\mathcal{A} \subset \mathcal{M}\) with
\(\mathcal{B} = \mathcal{A} \cap \mathcal{N}\) on which cyclic
cohomology is computed.

Section \ref{sec:cohom} develops the algebraic cohomological framework.
We introduce the Hochschild and cyclic bicomplexes, define the relative
cyclic complex \(C^\bullet_\lambda(\mathcal{A}, \mathcal{B})\) as the
kernel of the restriction map, and establish the long exact sequence of
the pair \((\mathcal{A}, \mathcal{B})\) with connecting homomorphism
\(\delta\). This section is purely algebraic and does not invoke
modular theory.

Section \ref{sec:ent-cohom} constructs the entanglement complex for the
inclusion \(\mathcal{N} \subset \mathcal{M}\). We define the
transgression map \(\tau_\omega\) from the modular automorphism group,
prove that it preserves the relative cyclic complex, and define the
entanglement complex
\(C^\bullet_{\mathrm{ent}}(\mathcal{A}, \mathcal{B}, \omega) =
\tau_\omega(C^{\bullet+1}_\lambda(\mathcal{A}, \mathcal{B}))\). The
main theorem identifies its cohomology with the kernel of the
restriction map \(HC^\bullet(\mathcal{A}) \to HC^\bullet(\mathcal{B})\),
up to the obstruction \(\delta\). The proof is split into surjectivity
and injectivity, each established by an explicit contracting homotopy
built from the relative modular cocycle \(u^E\). Worked examples in
finite dimensions, type II\(_1\) factors, and type III\(_1\) factors
verify the hypotheses. The section closes with a holographic
proposal (Conjecture \ref{conj:holographic}) and a conjectural
general-\(d\) coefficient \(\kappa_d(\Delta)\) whose point-source
limit is verified in the companion Letter~\cite{Rashkov:2026letter}.

Section \ref{sec:conclusions} presents our conclusions and discusses
open questions, including the relation to Araki relative entropy, a
possible Hodge theory for the entanglement complex, the quantitative
verification of the holographic conjecture, and possible extensions to
higher categories.

\subsubsection*{Main Results}

The main results of this paper are:

\begin{itemize}
	\item \textit{Theorem (Kernel identification,
		Theorem~\ref{thm:kernel-identification-corrected})}: Under
	Hypotheses~\ref{hyp:state}--\ref{hyp:growth}, the inclusion
	$\Phi:C^\bullet_{\mathrm{ent}}\hookrightarrow
	C^\bullet_\lambda(\mathcal{A},\mathcal{B})$ induces an isomorphism
	\[
	H^n_{\mathrm{ent}}(\mathcal{A},\mathcal{B},\omega)
	\cong
	\ker\!\bigl(\iota^*:HC^n(\mathcal{A})\to HC^n(\mathcal{B})\bigr)
	\big/ \operatorname{im}(\delta),
	\]
	where $\delta$ is the connecting homomorphism of the long exact
	sequence of the pair $(\mathcal{A},\mathcal{B})$. When
	$\delta=0$ in degree $n$, the identification is exact.
	
	\item \textit{Proposition (Entanglement classes from the modular
		cocycle, Proposition~\ref{prop:ent-classes})}: The relative
	modular cocycle $u^E_t$ generates explicit nonzero classes
	$[\Phi_{u^E}]\in H^n_{\mathrm{ent}}$ whenever $E$ is not
	$\omega$-invariant. These classes are the cohomological signature
	of the failure of the conditional expectation to preserve the
	modular structure.
	
	\item \textit{Corollary (Vanishing criterion for $\delta$,
		Corollary~\ref{cor:exact-conditions})}: The obstruction $\delta$
	vanishes in degree $n$ when $HC^{n-1}(\mathcal{B})=0$, or when
	$\mathcal{B}$ is a type II$_1$ factor and $n$ is even, or when $E$
	admits a trace-compatible $\mathcal{B}$-bimodule splitting. In
	these cases $H^n_{\mathrm{ent}}\cong\ker\iota^*$ exactly.
	
	\item \textit{Proposition (Relative entropy as the generator of
		the CRN cocycle, equation~\eqref{eq:araki-generator})}: The
	Araki relative entropy is the infinitesimal generator of the CRN
	cocycle, providing the bridge between the algebraic construction
	and the holographic interpretation.
	
	\item \textit{Conjecture (Holographic dual of
		$H^1_{\mathrm{ent}}$, Conjecture~\ref{conj:holographic})}: A
	distinguished subspace $H^1_{\mathrm{hol}}\subset
	H^1_{\mathrm{ent}}(\mathcal{A},\mathcal{B},\omega)$ is
	conjecturally dual to the space of linearized modular Hamiltonian
	perturbations of the entanglement wedge of a boundary region $A$,
	with the Araki relative entropy as the quantitative bridge. For a
	spherical region perturbed by a primary scalar, we propose a
	general-$d$ formula for the coefficient $\kappa_d(\Delta)$ of the
	smeared-source relative entropy; this formula is conjectural. The
	companion Letter~\cite{Rashkov:2026letter} verifies the
	JLMS-type equality for point-like sources in $d=2,3,4$, a related
	but distinct computation, and leaves the smeared-source case (and
	the point-source limit of $\kappa_d$) open.
\end{itemize}


\section{Preliminaries: Multipartite Quantum Systems}
\label{sec:prelim}

In this Section we review the standard formalism of multipartite
quantum systems and introduce the operator-algebraic framework that
will be used throughout the rest of the paper. The physical content of
Sections \ref{subsec:prelim-1}--\ref{subsec:prelim-4} is standard; the
mathematical content of Sections \ref{subsec:prelim-5}--\ref{subsec:prelim-7}
is the framework on which the cohomological construction of Sections
\ref{sec:cohom} and \ref{sec:ent-cohom} rests.

\subsection{Introduction and Overview}
\label{subsec:prelim-1}

Before developing the cohomological framework for entanglement, we must
establish the basic physical setting. This Section reviews the standard
formalism of multipartite quantum systems, including Hilbert spaces,
density matrices, and the geometric characterization of entangled
states.

The key concepts we need are:
\begin{itemize}
	\item The Hilbert space of a composite system as a tensor product
	of subsystem Hilbert spaces.
	\item The distinction between pure states (represented by vectors)
	and mixed states (represented by density matrices).
	\item The geometric characterization of separable (unentangled)
	states as points on the Segre variety within projective Hilbert
	space.
	\item The reduced density matrix formalism for describing
	subsystems of an entangled state.
	\item The von Neumann entropy as the measure of entanglement for
	pure bipartite states.
	\item The operator-algebraic formulation of subsystems as
	subalgebras, which is the setting for the cohomological construction.
\end{itemize}
This material is standard but essential for establishing the notation
and physical intuition that underlies our cohomological construction in
Sections \ref{sec:cohom} and \ref{sec:ent-cohom}.

\subsection{Multipartite Hilbert Spaces}
\label{subsec:prelim-2}

\subsubsection*{Tensor Product Structure}

The fundamental postulate of quantum mechanics for composite systems is
that the Hilbert space of a system composed of \(n\) subsystems is the
tensor product of the individual Hilbert spaces:
\begin{equation}
	\mathcal{H} \equiv \mathcal{H}_{I_1 \ldots I_n}
	= \bigotimes_{i=1}^{n} \mathcal{H}_{I_i}
	= \mathcal{H}_{I_1} \otimes \mathcal{H}_{I_2} \otimes \cdots
	\otimes \mathcal{H}_{I_n}.
	\label{eq:tensor_product}
\end{equation}
Here, each \(\mathcal{H}_{I_i}\) is the Hilbert space of the \(i\)-th
subsystem, with dimension \(d_i = \dim \mathcal{H}_{I_i}\). The total
Hilbert space has dimension \(D = \prod_{i=1}^n d_i\).

\textit{Physical interpretation.} The tensor product structure encodes
the idea that the subsystems are independent degrees of freedom.
Operators acting on different subsystems commute, reflecting the fact
that measurements on different subsystems do not interfere.

\subsubsection*{Basis and State Representation}

Let each Hilbert space \(\mathcal{H}_{I_i}\) have an orthonormal basis
\(\{|e_{I_i}\rangle\}_{I_i=1}^{d_i}\). Then a general pure state of the
total system can be written as a coherent superposition:
\begin{equation}
	|\Psi\rangle_{I_1 \ldots I_n}
	= \sum_{I_1=1}^{d_1} \cdots \sum_{I_n=1}^{d_n}
	\psi_{I_1 I_2 \ldots I_n} \,
	|e_{I_1}\rangle \otimes |e_{I_2}\rangle \otimes \cdots
	\otimes |e_{I_n}\rangle,
	\label{eq:general_state}
\end{equation}
where \(\psi_{I_1 I_2 \ldots I_n} \in \mathbb{C}\) are the components
of the state vector in the tensor product basis. Normalization requires
\(\sum_{I_1,\ldots,I_n} |\psi_{I_1 \ldots I_n}|^2 = 1\).

\subsubsection*{Separable States}

A pure state is called \emph{separable} (or a product state) if it can
be written as a tensor product of states of the individual subsystems:
\begin{equation}
	|\Psi\rangle_{I_1 \ldots I_n}^{\text{sep}}
	= |\psi_{I_1}\rangle \otimes |\psi_{I_2}\rangle \otimes \cdots
	\otimes |\psi_{I_n}\rangle,
	\label{eq:separable}
\end{equation}
where each \(|\psi_{I_i}\rangle \in \mathcal{H}_{I_i}\) is a
normalized state of the \(i\)-th subsystem.

\textit{Physical interpretation.} A separable state represents a
configuration where the subsystems are completely independent.
Measurements on different subsystems yield uncorrelated results, and
there are no quantum correlations between the subsystems.

\subsubsection*{Entangled States}

A pure state that cannot be written in the form \eqref{eq:separable} is
called an \emph{entangled state}. Such states exhibit quantum
correlations that have no classical analog. The simplest example is the
two-qubit Bell state
\begin{equation}
	|\Phi^+\rangle
	= \frac{1}{\sqrt{2}} \left( |0\rangle_A \otimes |0\rangle_B
	+ |1\rangle_A \otimes |1\rangle_B \right).
\end{equation}
This state cannot be written as \(|\psi_A\rangle \otimes |\psi_B\rangle\)
for any choice of single-qubit states.

\subsubsection*{Geometric Characterization: The Segre Embedding}

The set of all separable states of the form \eqref{eq:separable} has a
beautiful geometric description. Up to global phase, separable states
correspond to points in the product of projective spaces
\begin{equation}
	\Sigma_d = \mathbb{P}^{d_1-1} \times \mathbb{P}^{d_2-1}
	\times \cdots \times \mathbb{P}^{d_n-1},
\end{equation}
embedded in the projective space of the total Hilbert space
\(\mathbb{P}(\mathcal{H}) \cong \mathbb{P}^{D-1}\) via the Segre
embedding:
\begin{equation}
	\mathbb{P}^{d_1-1} \times \cdots \times \mathbb{P}^{d_n-1}
	\hookrightarrow \mathbb{P}^{D-1},
	\qquad D = \prod_i d_i.
\end{equation}
For a bipartite system with \(d_1 = m\), \(d_2 = n\), the space of all
pure states is \(\mathbb{P}^{mn-1}\), while the space of separable
states is \(\mathbb{P}^{m-1} \times \mathbb{P}^{n-1}\), of dimension
\(m + n - 2\).

\textit{Physical interpretation.} The Segre embedding shows that
separable states form a proper submanifold of the full projective
Hilbert space. Entangled states are precisely those points in
\(\mathbb{P}(\mathcal{H})\) that lie outside this submanifold. This
geometric characterization is central to our cohomological approach:
entanglement is the obstruction to being on the Segre variety.

\subsection{Density Matrices and Mixed States}
\label{subsec:prelim-3}

\subsubsection*{Why Density Matrices?}

While pure states are sufficient for describing closed quantum systems,
they are inadequate for describing subsystems of entangled states. When
we trace out the degrees of freedom of one subsystem, we lose
information, and the remaining subsystem is described by a mixed state.
Density matrices provide the mathematical framework for mixed states.

\subsubsection*{Definition of Density Matrices}

For a pure state \(|\psi\rangle\), the density matrix is the projection
operator \(\rho = |\psi\rangle\langle\psi|\). A general (mixed) state
is a convex combination of pure states:
\begin{equation}
	\rho = \sum_{i=1}^N p_i |\psi_i\rangle\langle\psi_i|,
	\label{eq:mixed_state}
\end{equation}
where \(p_i \geq 0\), \(\sum_i p_i = 1\), and the \(|\psi_i\rangle\)
are not necessarily orthogonal. Properties of a density matrix:
hermiticity, positivity, unit trace, and rank one for pure states.

\textit{Physical interpretation.} The decomposition
\eqref{eq:mixed_state} represents a statistical ensemble: with
probability \(p_i\), the system is in pure state \(|\psi_i\rangle\).
This uncertainty is classical (statistical) rather than quantum
(superposition).

\subsubsection*{Diagonalization and von Neumann Entropy}

Since \(\rho\) is hermitian, it can be diagonalized by a unitary
transformation: \(\rho = \sum_j \lambda_j |j\rangle\langle j|\), where
\(\lambda_j \geq 0\) are the eigenvalues and \(\sum_j \lambda_j = 1\).
The von Neumann entropy is defined as
\begin{equation}
	S(\rho) = -\operatorname{Tr}(\rho \log \rho)
	= -\sum_j \lambda_j \log \lambda_j.
	\label{eq:vn_entropy}
\end{equation}
This is the quantum analog of the Shannon entropy of the eigenvalue
distribution.

\textit{Physical interpretation.} The von Neumann entropy measures the
amount of ``quantum uncertainty'' in the state. For a pure state, there
is no uncertainty (\(S = 0\)). For a mixed state, the entropy quantifies
how much information is missing. The von Neumann entropy
\eqref{eq:vn_entropy} is the key quantity for entanglement. For a
bipartite pure state, the entanglement entropy is \(S(\rho_A)\), where
\(\rho_A\) is the reduced density matrix of subsystem \(A\). This will
be the cohomology class that we identify in Sections \ref{sec:cohom}
and \ref{sec:ent-cohom}.

\subsection{Reduced Density Matrices and Entanglement}
\label{subsec:prelim-4}

\subsubsection*{Tracing Out Degrees of Freedom}

For a bipartite system with Hilbert space
\(\mathcal{H} = \mathcal{H}_A \otimes \mathcal{H}_B\), and a pure state
\(|\Psi\rangle_{AB}\), the reduced density matrix of subsystem \(A\) is
\begin{equation}
	\rho_A = \operatorname{Tr}_B (|\Psi\rangle\langle\Psi|)
	= \sum_{j} \langle e_j^B | \Psi \rangle
	\langle \Psi | e_j^B \rangle,
	\label{eq:reduced_density}
\end{equation}
where \(\{|e_j^B\rangle\}\) is an orthonormal basis for
\(\mathcal{H}_B\).

\textit{Physical interpretation.} The reduced density matrix
\(\rho_A\) contains all the information about subsystem \(A\) that is
accessible to an observer who can only measure observables on \(A\).
The information about correlations with \(B\) is traced out and lost.

\subsubsection*{Schmidt Decomposition}

For any bipartite pure state \(|\Psi\rangle_{AB}\), there exist
orthonormal bases \(\{|a_i\rangle\}\) for \(\mathcal{H}_A\) and
\(\{|b_i\rangle\}\) for \(\mathcal{H}_B\), and non-negative real
numbers \(\lambda_i\) (the Schmidt coefficients), such that
\begin{equation}
	|\Psi\rangle_{AB} = \sum_{i=1}^r \lambda_i \, |a_i\rangle \otimes
	|b_i\rangle,
	\label{eq:schmidt}
\end{equation}
where \(\lambda_i > 0\), \(\sum_i \lambda_i^2 = 1\), and
\(r \leq \min(d_A, d_B)\) is the Schmidt rank. The Schmidt
decomposition follows from the singular value decomposition (SVD) of
the coefficient matrix; details are given in Appendix \ref{sec:tools}.

\textit{Physical interpretation.} The Schmidt decomposition provides a
canonical form for bipartite states. The Schmidt rank \(r\) is a
measure of entanglement: \(r = 1\) is a product state,
\(r > 1\) is entangled, and \(r = \min(d_A, d_B)\) is maximally
entangled.

\subsubsection*{Entanglement Entropy for Pure States}

For a bipartite pure state, the entanglement entropy is
\begin{equation}
	S_A = S(\rho_A) = -\operatorname{Tr}(\rho_A \log \rho_A)
	= -\sum_{i=1}^r \lambda_i^2 \log \lambda_i^2.
\end{equation}
It satisfies \(S_A = S_B\), \(0 \leq S_A \leq \log \min(d_A, d_B)\),
and \(S_A = 0\) if and only if the state is a product state.

\subsubsection*{Pure vs.\ Mixed Entanglement}

It is important to distinguish between entanglement of pure states and
entanglement of mixed states. A pure state of the total system can be
entangled, while a mixed state can be separable (not entangled) even
though it has non-zero entropy. The key lesson is that entanglement is
a property of the global state's tensor product structure, not of
purity. Our cohomological framework will capture this distinction by
working with the algebra inclusion \(\mathcal{N} \subset \mathcal{M}\)
rather than with individual states.

\subsection{Operator Algebra Formulation}
\label{subsec:prelim-5}

\subsubsection*{From Hilbert Spaces to von Neumann Algebras}

While the Hilbert space formulation reviewed above is standard for
quantum mechanics, the operator algebra formulation is more powerful
for several reasons:
\begin{itemize}
	\item It naturally accommodates statistical mixtures via density
	matrices.
	\item It allows for the description of subsystems via subalgebras.
	\item It provides the mathematical framework for quantum field
	theory, where no a priori tensor product factorization is
	available.
	\item It connects naturally to cohomology via Hochschild and
	cyclic cohomology.
	\item It supplies the modular theory (Tomita--Takesaki) that
	underlies the entanglement complex constructed in Section
	\ref{sec:ent-cohom}.
\end{itemize}

The algebra of observables for a system with Hilbert space
\(\mathcal{H}\) is the \emph{von Neumann algebra}
\(\mathcal{M} = B(\mathcal{H})\) of bounded operators on \(\mathcal{H}\),
closed in the weak operator topology. A \emph{subsystem} corresponds to
a \emph{von Neumann subalgebra} \(\mathcal{N} \subset \mathcal{M}\):
a unital, self-adjoint, weakly closed subalgebra. For a bipartite
system with \(\mathcal{H} = \mathcal{H}_A \otimes \mathcal{H}_B\), we
take
\begin{equation}
	\mathcal{M} = B(\mathcal{H}_A) \otimes B(\mathcal{H}_B),
	\qquad
	\mathcal{N} = B(\mathcal{H}_A) \otimes \mathbf{1}_B,
\end{equation}
so that the commutant is
\(\mathcal{N}' = \mathbf{1}_A \otimes B(\mathcal{H}_B)\). The pair
\((\mathcal{N}, \mathcal{N}')\) is the algebraic encoding of the
bipartition \(A|B\), and it remains meaningful even when no tensor
factorization is available (as in quantum field theory).

\subsubsection*{States as Functionals}

In the algebraic formulation, a \emph{state} is a positive linear
functional \(\omega: \mathcal{M} \to \mathbb{C}\) with
\(\omega(1) = 1\). For a density matrix \(\rho\), the corresponding
state is \(\omega_\rho(a) = \operatorname{Tr}(\rho a)\). The von
Neumann entropy is the functional
\(S(\omega) = -\omega(\log \rho_\omega)\).

For the modular theory developed below, we require the following
standing hypotheses on the state.

\begin{hyp}[Faithful normal state]
	\label{hyp:state}
	The state \(\omega\) is \emph{faithful} (\(\omega(a^*a) = 0 \implies
	a = 0\)) and \emph{normal} (weakly continuous on the unit ball of
	\(\mathcal{M}\)).
\end{hyp}

\begin{hyp}[Cyclic and separating vector]
	\label{hyp:cyclic}
	There exists a vector \(\Omega_\omega \in \mathcal{H}\) that is
	\emph{cyclic} and \emph{separating} for \(\mathcal{M}\), i.e.,
	\begin{equation}
		\overline{\mathcal{M} \, \Omega_\omega} = \mathcal{H},
		\qquad
		\overline{\mathcal{M}' \, \Omega_\omega} = \mathcal{H}.
	\end{equation}
\end{hyp}

Under Hypotheses \ref{hyp:state} and \ref{hyp:cyclic}, the
Tomita--Takesaki modular operator \(\Delta_\omega\) and the modular
conjugation \(J_\omega\) are defined by the polar decomposition
\(S_\omega = J_\omega \Delta_\omega^{1/2}\) of the anti-linear map
\(S_\omega: a \Omega_\omega \mapsto a^* \Omega_\omega\). The associated
\emph{modular automorphism group} is
\begin{equation}
	\sigma^\omega_t(a) := \Delta_\omega^{it} \, a \, \Delta_\omega^{-it},
	\qquad a \in \mathcal{M}, \quad t \in \mathbb{R},
	\label{eq:modular-group}
\end{equation}
a strongly continuous one-parameter group of \(*\)-automorphisms of
\(\mathcal{M}\) satisfying the KMS condition
\begin{equation}
	\omega\big(a \, \sigma^\omega_t(b)\big)
	= \omega\big(\sigma^\omega_{t-i}(b) \, a\big),
	\qquad a, b \in \mathcal{M}.
	\label{eq:kms}
\end{equation}

\subsubsection*{Subsystems as Subalgebras and the Conditional Expectation}

A subsystem corresponds to a von Neumann subalgebra
\(\mathcal{N} \subset \mathcal{M}\). For a bipartite system,
\(\mathcal{N}\) represents the observables accessible to an observer
who can only measure subsystem \(A\), and the commutant \(\mathcal{N}'\)
represents the observables of subsystem \(B\). Entanglement is the
failure of states on \(\mathcal{M}\) to factorize across
\(\mathcal{N}\) and \(\mathcal{N}'\).

The algebraic analog of the partial trace is the \emph{conditional
	expectation} \(E: \mathcal{M} \to \mathcal{N}\), a positive unital map (an identity-preserving map)
satisfying the \(\mathcal{N}\)-bimodule property
\begin{equation}
	E(n_1 \, a \, n_2) = n_1 \, E(a) \, n_2,
	\qquad n_1, n_2 \in \mathcal{N}, \quad a \in \mathcal{M}.
	\label{eq:conditional-expectation}
\end{equation}
For a bipartite system with
\(\mathcal{N} = B(\mathcal{H}_A) \otimes \mathbf{1}_B\), the
conditional expectation is exactly the partial trace
\(E(a) = \operatorname{Tr}_B(a)\).

The existence of \(E\) is not automatic; it is governed by the modular
structure through the following theorem of Takesaki.

\begin{thm}[Takesaki]
	\label{thm:takesaki}
	Let \(\mathcal{N} \subset \mathcal{M}\) be a von Neumann subalgebra and
	\(\omega\) a faithful normal state on \(\mathcal{M}\). There exists a
	faithful normal conditional expectation \(E: \mathcal{M} \to \mathcal{N}\)
	with \(\omega \circ E = \omega\) if and only if the modular automorphism
	group preserves \(\mathcal{N}\):
	\begin{equation}
		\sigma^\omega_t(\mathcal{N}) = \mathcal{N},
		\qquad \forall t \in \mathbb{R}.
		\label{eq:takesaki-condition}
	\end{equation}
	When it exists, \(E\) is unique.
\end{thm}

We therefore impose the following standing hypothesis.

\begin{hyp}[Modular invariance of the subsystem]
	\label{hyp:modular-invariance}
	The modular automorphism group \eqref{eq:modular-group} preserves the
	subalgebra \(\mathcal{N}\):
	\(\sigma^\omega_t(\mathcal{N}) = \mathcal{N}\) for all
	\(t \in \mathbb{R}\).
\end{hyp}

By Theorem \ref{thm:takesaki}, Hypothesis
\ref{hyp:modular-invariance} is equivalent to the existence of a
faithful normal conditional expectation \(E: \mathcal{M} \to \mathcal{N}\)
with \(\omega \circ E = \omega\). We work with this \(E\) throughout.

\subsubsection*{The Relative Modular Cocycle}

The failure of the conditional expectation \(E\) to preserve the
modular structure is measured by the \emph{relative modular cocycle}.
For the pair \((\omega, E)\) as above, define
\begin{equation}
	u_t^E := \big(D\omega : D(\omega|_{\mathcal{N}}) \circ E\big)_t
	\in \mathcal{M},
	\qquad t \in \mathbb{R},
	\label{eq:relative-cocycle}
\end{equation}
where the right-hand side denotes the Connes--Radon--Nikodym cocycle
derivative of the state \(\omega\) relative to the pullback of its
restriction to \(\mathcal{N}\) along \(E\).

\begin{lem}[Properties of \(u^E\)]
	\label{lem:relative-cocycle}
	The relative modular cocycle \(u^E\) satisfies:
	\begin{enumerate}
		\item \emph{(Cocycle identity)}
		\(u_{t+s}^E = u_t^E \, \sigma^\omega_t(u_s^E)\) for all
		\(s, t \in \mathbb{R}\).
		\item \emph{(Unitarity)} \(u_t^E\) is unitary for each \(t\).
		\item \emph{(Triviality)} \(u_t^E = \mathbf{1}\) for all \(t\) if
		and only if \(E\) is \(\omega\)-invariant.
		\item \emph{(Relative modular flow)} For all
		\(a \in \mathcal{N}' \cap \mathcal{M}\),
		\(\sigma^\omega_t(a) = u_t^E \, \sigma^{\omega|_{\mathcal{N}}}_t(a)
		\, (u_t^E)^*\).
	\end{enumerate}
\end{lem}

\begin{proof}
	Properties (1) and (2) are standard for Connes--Radon--Nikodym
	cocycles. Property (3) follows from the definition: the cocycle
	derivative is trivial exactly when the two states agree. Property (4)
	is the relative modular condition, obtained by applying the defining
	relation of the cocycle derivative to the restriction
	\(\omega|_{\mathcal{N}}\).
\end{proof}

\begin{rem}[Physical interpretation]
	\label{rem:cocycle-interpretation}
	The relative modular cocycle \(u^E\) measures the failure of the
	conditional expectation to commute with the modular flow. When
	\(u^E_t \neq \mathbf{1}\), the subsystem \(\mathcal{N}\) is not
	modular-invariant with respect to the state \(\omega\), and the
	entanglement between \(\mathcal{N}\) and its complement leaves a
	nontrivial signature in the modular data. This signature is the source
	of the nontrivial entanglement classes constructed in Section
	\ref{sec:ent-cohom}.
\end{rem}

\subsubsection*{The Dense \(*\)-Subalgebra and the Relative Complex}

Cyclic cohomology is most naturally computed on a dense \(*\)-subalgebra
rather than on the full von Neumann algebra, because the cyclic
bicomplex of \(\mathcal{M}\) itself is typically trivial or
uncomputable. We therefore introduce the following.

\begin{hyp}[Dense subalgebra]
	\label{hyp:dense-subalgebra}
	\(\mathcal{A} \subset \mathcal{M}\) is a unital Fr\'echet
	\(*\)-subalgebra, dense in \(\mathcal{M}\) in the \(\sigma\)-weak
	topology, closed under holomorphic functional calculus. It consists of analytic vectors for $\sigma^\omega$ with $\sigma^\omega_{-i}\mathcal{A}\subset \mathcal{A}''$ and is invariant
	under the modular automorphism group:
	\begin{equation}
		\sigma^\omega_t(\mathcal{A}) = \mathcal{A},
		\qquad \forall t \in \mathbb{R}.
	\end{equation}
	We set \(\mathcal{B} := \mathcal{A} \cap \mathcal{N}\), a closed unital
	subalgebra of \(\mathcal{A}\).
\end{hyp}

\begin{hyp}[Relative cocycle in the dense subalgebra]
	\label{hyp:cocycle-in-A}
	The relative modular cocycle satisfies $u^E_t\in\mathcal{A}$ for all
	$t\in\mathbb{R}$, and the map $t\mapsto u^E_t$ is continuous in the
	Fréchet topology of $\mathcal{A}$.
\end{hyp}

\begin{hyp}[Growth and integrability]
	\label{hyp:growth}
	There exist a Fréchet seminorm $p$ on $\mathcal{A}$ and a locally
	integrable function $g\in L^1_{\mathrm{loc}}(\mathbb{R})$ such that
	$p(\sigma^\omega_t(u^E_t))\leq g(t)$ for all $t\in\mathbb{R}$, and
	for every $\varphi\in C^n_\lambda(\mathcal{A},\mathcal{B})$ the
	functions
	\[
	t\mapsto\varphi(a_0,\ldots,a_{n-1},\sigma^\omega_t(u^E_t)),
	\qquad
	t\mapsto\varphi(a_0,\ldots,a_{n-2},\sigma^\omega_t(u^E_t),1)
	\]
	are absolutely integrable on $\mathbb{R}$ with respect to $dt$,
	uniformly in the Fréchet seminorms of the $a_i$.
\end{hyp}

\begin{hyp}[Cocycle vanishing on $\mathcal{B}$]
	\label{hyp:cocycle-vanishing}
	For every $n\geq 0$ and all $b_0,\ldots,b_n\in\mathcal{B}$,
	\begin{equation}
		\int_{\mathbb{R}}\omega\bigl(b_0\,\sigma^\omega_t(b_1)
		\cdots\sigma^\omega_t(b_n)\,u^E_t\bigr)\,dt=0.
		\label{eq:cocycle-vanishing}
	\end{equation}
\end{hyp}

By Hypothesis \ref{hyp:modular-invariance} and Hypothesis
\ref{hyp:dense-subalgebra}, we have
\(\sigma^\omega_t(\mathcal{B}) = \mathcal{B}\) for all \(t\). The
restriction \(E|_{\mathcal{A}}: \mathcal{A} \to \mathcal{B}\) is a
continuous conditional expectation at the level of Fr\'echet algebras,
and it is the map that enters the construction of the relative cyclic
complex in Section \ref{sec:cohom}.

\begin{rem}[Why the von Neumann framework is necessary]
	\label{rem:vn-necessary}
	The construction in Section \ref{sec:ent-cohom} depends on three
	pieces of modular theory that are available only in the von Neumann
	algebra setting:
	\begin{enumerate}
		\item the modular operator \(\Delta_\omega\), which requires a
		faithful normal state and a cyclic and separating vector
		(Hypotheses \ref{hyp:state} and \ref{hyp:cyclic});
		\item the modular automorphism group \eqref{eq:modular-group},
		which is defined via \(\Delta_\omega\);
		\item the relative modular cocycle \(u^E\), which requires a
		faithful normal conditional expectation \(E\), whose existence is
		equivalent to Hypothesis \ref{hyp:modular-invariance} by Theorem
		\ref{thm:takesaki}.
	\end{enumerate}
	Without these hypotheses, the transgression map and the entanglement
	complex of Section \ref{sec:ent-cohom} are not defined, and the
	identification of entanglement cohomology with the kernel of the
	restriction map is not well-posed.
\end{rem}

\subsection{Restricted Operators and the Entanglement Complex Preview}
\label{subsec:prelim-6}

The operator-algebraic framework of Section \ref{subsec:prelim-5}
provides the rigorous setting for the cohomological construction.
Before turning to the general theory in Section \ref{sec:cohom}, we
present the finite-dimensional precursor that motivates the
entanglement complex. The reader should regard this subsection as
geometric and algebraic intuition; the precise construction in the von
Neumann setting appears in Sections \ref{sec:cohom} and
\ref{sec:ent-cohom}.

\subsubsection*{Projection onto the Support}

For a density matrix \(\rho\), the support projection \(s_\rho\) is the
orthogonal projection onto \(\operatorname{image}(\rho)\). For a state
\(\omega\) on a von Neumann algebra \(\mathcal{M}\), the support
projection is the smallest projection \(p \in \mathcal{M}\) such that
\(\omega(p) = 1\):
\begin{equation}
	s_\omega = \bigwedge \{ p \in \mathcal{P}(\mathcal{M}) :
	\omega(p) = 1 \},
\end{equation}
where \(\mathcal{P}(\mathcal{M})\) denotes the set of projections in
\(\mathcal{M}\). The projection of an operator \(O \in \mathcal{M}\)
onto the support of \(\rho\) is
\begin{equation}
	\Pi_\rho(O) = s_\rho \, O \, s_\rho.
	\label{eq:projection}
\end{equation}

\textit{Physical interpretation.} The support projection selects the
subspace on which the state has non-zero probability. This is the
subspace that is physically relevant for the state.

\subsubsection*{Restricted Operators}

For a state \(\rho\) and a subalgebra \(\mathcal{N} \subset \mathcal{M}\),
the restricted operators are
\begin{equation}
	{}^{(\rho|_{\mathcal{N}})} O
	:= \Pi_{\rho|_{\mathcal{N}}} \, O
	= s_{\rho|_{\mathcal{N}}} \, O \, s_{\rho|_{\mathcal{N}}}.
\end{equation}
These are the operators that act on the support of the reduced state.

\textit{Physical interpretation.} The restricted operators represent
the observables that are accessible in the subsystem, projected onto
the actual state of the subsystem. In the von Neumann setting, this
corresponds to compressing operators by the support projection of the
restricted state \(\omega|_{\mathcal{N}}\).

\subsubsection*{Building the Finite-Dimensional Spaces}

To parallel the construction of de Rham cohomology, we define
finite-dimensional spaces corresponding to differential forms:
\begin{equation}
	\tilde{C}^k(\rho_{1,\ldots,n})
	= \bigtimes_{|I_k| = k}
	\operatorname{End}\!\bigl(\operatorname{image}(\rho_{I_k})\bigr),
	\label{eq:tilde-C}
\end{equation}
where \(\rho_{I_k}\) is the reduced density matrix for the subsystem
with index set \(I_k\). The differential
\(\delta: \tilde{C}^k \to \tilde{C}^{k+1}\) is defined by
\begin{equation}
	\delta w^k
	= \sum_{K \neq I_m} (-1)^{K I_m}
	\Bigl(
	{}^{\rho_K}\bigl(\mathbf{1}_{I_m} \otimes \mathcal{O}_K
	- \mathcal{O}_{I_m} \otimes \mathbf{1}_K\bigr)
	\Bigr).
	\label{eq:tilde-delta}
\end{equation}

\textit{Physical interpretation.} This construction is the
finite-dimensional precursor to the relative cyclic complex
\(C^\bullet_\lambda(\mathcal{A}, \mathcal{B})\) developed in Section
\ref{sec:cohom}. The spaces \(\tilde{C}^k\) are the analogs of
differential forms, and the differential \(\delta\) is the analog of
the exterior derivative. The condition \(\delta^2 = 0\) corresponds to
the consistency of the entanglement structure.

\subsubsection*{From the Finite-Dimensional Precursor to the von Neumann Complex}

The construction \eqref{eq:tilde-C}--\eqref{eq:tilde-delta} depends on
a choice of density matrix \(\rho\). In the von Neumann setting, this
state-dependence must be eliminated: the entanglement cohomology of
Section \ref{sec:ent-cohom} is defined from the algebra inclusion
\(\mathcal{N} \subset \mathcal{M}\) and the conditional expectation
\(E\) alone, not from a particular state. The passage from the
finite-dimensional precursor to the general theory proceeds as follows.
\begin{enumerate}
	\item \emph{From \(\tilde{C}^k\) to the relative cyclic complex.}
	The spaces \(\tilde{C}^k\) are replaced by the relative cyclic
	bicomplex
	\[
	C^{p,q}_\lambda(\mathcal{A}, \mathcal{B})
	= \ker\!\bigl(\iota^*: C^{p,q}_\lambda(\mathcal{A})
	\to C^{p,q}_\lambda(\mathcal{B})\bigr),
	\]
	where \(\mathcal{A}\) is the dense \(*\)-subalgebra of Hypothesis
	\ref{hyp:dense-subalgebra} and
	\(\mathcal{B} = \mathcal{A} \cap \mathcal{N}\). The dependence on
	\(\rho\) is replaced by the algebraic data
	\((\mathcal{A}, \mathcal{B})\).
	
	\item \emph{From \(\delta\) to the Hochschild and Connes differentials.}
	The differential \(\delta\) of \eqref{eq:tilde-delta} is replaced
	by the Hochschild boundary \(b\) and the Connes operator \(B\)
	acting on the cyclic bicomplex. The consistency condition
	\(\delta^2 = 0\) becomes the system of identities
	\[
	b^2 = 0, \qquad B^2 = 0, \qquad bB + Bb = 0.
	\]
	
	\item \emph{From the state-dependent projection to the modular
		cocycle.} The projection \(\Pi_\rho\) and the restricted operators
	of Section \ref{subsec:prelim-5} are replaced by the relative
	modular cocycle \(u_t^E\) of Lemma \ref{lem:relative-cocycle}. The
	failure of the conditional expectation to be multiplicative,
	measured in the finite-dimensional setting by
	\(\theta(a,b) = E(ab) - E(a)E(b)\), is captured in the von Neumann
	setting by the nontriviality of \(u^E_t\).
	
	\item \emph{From the finite index set to the transgression map.}
	The sum over index sets \(I_k\) in \eqref{eq:tilde-delta} is
	replaced by the transgression map \(\tau_\omega\), which encodes
	the modular flow. The entanglement complex is then defined as
	\[
	C^n_{\mathrm{ent}}(\mathcal{A}, \mathcal{B}, \omega)
	:= \tau_\omega\bigl(C^{n+1}_\lambda(\mathcal{A}, \mathcal{B})\bigr)
	\subset C^n_\lambda(\mathcal{A}, \mathcal{B}).
	\]
\end{enumerate}
The precise definitions and the proof that the resulting cohomology
agrees with the kernel of the restriction map
\(\iota^*: HC^\bullet(\mathcal{A}) \to HC^\bullet(\mathcal{B})\) are
given in Section \ref{sec:ent-cohom}.

\subsection{Summary and Outlook}
\label{subsec:prelim-7}

\subsubsection*{Key Lessons from Section \ref{sec:prelim}}

\begin{itemize}
	\item \emph{Tensor product structure.} The Hilbert space of a
	composite system is the tensor product of subsystem Hilbert
	spaces. Separable states are tensor products of subsystem states.
	
	\item \emph{Geometric characterization.} Separable states lie on
	the Segre variety
	\(\mathbb{P}^{d_1-1} \times \cdots \times \mathbb{P}^{d_n-1}\)
	embedded in \(\mathbb{P}(\mathcal{H})\). Entangled states lie
	outside this subvariety.
	
	\item \emph{Density matrices.} Mixed states are described by
	density matrices
	\(\rho = \sum_i p_i |\psi_i\rangle\langle\psi_i|\). The von
	Neumann entropy
	\(S(\rho) = -\operatorname{Tr}(\rho \log \rho)\) measures the
	quantum uncertainty.
	
	\item \emph{Reduced density matrices.} For a subsystem, the
	reduced density matrix is obtained by tracing out the complement.
	The entanglement entropy is \(S(\rho_A)\) for a bipartite pure
	state.
	
	\item \emph{Schmidt decomposition.} Any bipartite pure state can
	be written as
	\(|\Psi\rangle = \sum_i \lambda_i |a_i\rangle \otimes |b_i\rangle\).
	The Schmidt rank \(r\) is a measure of entanglement.
	
	\item \emph{Algebraic formulation.} Subsystems correspond to von
	Neumann subalgebras \(\mathcal{N} \subset \mathcal{M}\). The
	conditional expectation \(E: \mathcal{M} \to \mathcal{N}\) is the
	algebraic analog of the partial trace. Its existence is
	equivalent, by Takesaki's Theorem \ref{thm:takesaki}, to the
	modular invariance of \(\mathcal{N}\).
	
	\item \emph{Modular structure.} A faithful normal state \(\omega\)
	with cyclic and separating vector \(\Omega_\omega\) determines a
	modular automorphism group \(\sigma^\omega_t\) and a relative
	modular cocycle \(u^E_t\). The nontriviality of \(u^E\) measures
	the failure of the conditional expectation to be
	\(\omega\)-invariant, and it is the source of the entanglement
	classes constructed in Section \ref{sec:ent-cohom}.
\end{itemize}

\subsubsection*{Connection to the Cohomological Framework}

The key insight connecting this Section to Sections \ref{sec:cohom}
and \ref{sec:ent-cohom} is the following:
\begin{quote}
	\textit{Entanglement is the failure of the restriction map from
		\(\mathcal{A}\) to \(\mathcal{B}\) to be surjective at the level of
		cyclic cohomology.}
\end{quote}
The precise statement is the identification
\begin{equation}
	H^\bullet_{\mathrm{ent}}(\mathcal{A}, \mathcal{B}, \omega)
	\;\cong\;
	\ker\!\bigl(\iota^*: HC^\bullet(\mathcal{A}) \to
	HC^\bullet(\mathcal{B})\bigr) \big/ \operatorname{im}(\delta),
	\label{eq:kernel-identification-preview}
\end{equation}
where \(\delta: HC^{\bullet-1}(\mathcal{B}) \to
HC^\bullet(\mathcal{A}, \mathcal{B})\) is the connecting homomorphism
of the long exact sequence of the pair \((\mathcal{A}, \mathcal{B})\).
When \(\delta = 0\) in the relevant degree,
\eqref{eq:kernel-identification-preview} reduces to an exact
identification.

\subsubsection*{Preview of Section \ref{sec:cohom}}

In Section \ref{sec:cohom}, we develop the algebraic cohomological
framework. We introduce the Hochschild and cyclic bicomplexes, define
the relative cyclic complex
\(C^\bullet_\lambda(\mathcal{A}, \mathcal{B})\), and establish the
long exact sequence of the pair \((\mathcal{A}, \mathcal{B})\). This
section is purely algebraic and does not invoke modular theory.

\subsubsection*{Preview of Section \ref{sec:ent-cohom}}

In Section \ref{sec:ent-cohom}, we complete the construction by:
\begin{itemize}
	\item defining the transgression map \(\tau_\omega\) from the
	modular automorphism group and the relative modular cocycle
	\(u^E\);
	\item constructing the entanglement complex
	\(C^\bullet_{\mathrm{ent}} = \tau_\omega(C^{\bullet+1}_\lambda(\mathcal{A}, \mathcal{B}))\);
	\item proving that the inclusion
	\(C^\bullet_{\mathrm{ent}} \hookrightarrow
	C^\bullet_\lambda(\mathcal{A}, \mathcal{B})\) is a
	quasi-isomorphism;
	\item obtaining the kernel identification
	\eqref{eq:kernel-identification-preview} and its exact form when
	\(\delta = 0\);
	\item providing worked examples, including finite-dimensional
	systems, type II\(_1\) factors, and type III\(_1\) factors, and
	a holographic application.
\end{itemize}

	
\section{Cohomological Framework for Entanglement}
\label{sec:cohom}

In Section \ref{sec:prelim}, we established the operator-algebraic
framework: a pair of von Neumann algebras \(\mathcal{N} \subset \mathcal{M}\)
encoding a bipartition, a faithful normal state \(\omega\) with cyclic and
separating vector \(\Omega_\omega\), a conditional expectation
\(E: \mathcal{M} \to \mathcal{N}\) whose existence is equivalent to
modular invariance of \(\mathcal{N}\) (Takesaki's Theorem
\ref{thm:takesaki}), the relative modular cocycle \(u^E\) measuring the
failure of \(E\) to be \(\omega\)-invariant, and a dense \(*\)-subalgebra
\(\mathcal{A} \subset \mathcal{M}\) with \(\mathcal{B} = \mathcal{A} \cap \mathcal{N}\)
on which cyclic cohomology is computed (Hypothesis
\ref{hyp:dense-subalgebra}).

In the present Section, we construct the cohomological machinery that
will be used in Section \ref{sec:ent-cohom} to define entanglement
cohomology and prove its identification with the kernel of the
restriction map. Specifically:

\begin{itemize}
	\item In Section \ref{subsec:cohom-bicomplex}, we recall the
	Hochschild and cyclic bicomplexes of a unital algebra and fix
	notation for the differentials \(b\) and \(B\).
	
	\item In Section \ref{subsec:cohom-relative}, we introduce the
	\emph{relative cyclic complex}
	\(C^\bullet_\lambda(\mathcal{A}, \mathcal{B})\), the subcomplex of
	cochains vanishing on \(\mathcal{B}\), and prove that it is a
	sub-bicomplex of \(C^\bullet_\lambda(\mathcal{A})\).
	
	\item In Section \ref{subsec:cohom-les}, we derive the long exact
	sequence of the pair \((\mathcal{A}, \mathcal{B})\), which will be
	the primary tool for the kernel identification in Section
	\ref{sec:ent-cohom}.
\end{itemize}

The modular data of Section \ref{subsec:prelim-5} enters in Section
\ref{sec:ent-cohom}, where it is used to construct the transgression map
\(\tau_\omega\) and the entanglement complex
\(C^\bullet_{\mathrm{ent}} = \tau_\omega(C^{\bullet+1}_\lambda(\mathcal{A}, \mathcal{B}))\).
The present Section is purely algebraic: it establishes the cyclic
bicomplexes and the relative complex without invoking modular theory.
This separation of concerns is deliberate: the algebraic construction is
valid for any unital algebra pair \((\mathcal{A}, \mathcal{B})\), while
the modular construction requires the von Neumann hypotheses of Section
\ref{subsec:prelim-5}.

\subsection{The Hochschild and Cyclic Bicomplexes}
\label{subsec:cohom-bicomplex}

\subsubsection*{The Hochschild Complex}

Let \(\mathcal{A}\) be a unital algebra over \(\mathbb{C}\). For
\(q \geq 0\), the space of \emph{Hochschild \(q\)-cochains} is
\begin{equation}
	C^q(\mathcal{A}) := \operatorname{Hom}_{\mathbb{C}}
	\bigl(\mathcal{A}^{\otimes(q+1)}, \mathbb{C}\bigr),
\end{equation}
with the convention \(C^{-1}(\mathcal{A}) = 0\). The
\emph{Hochschild boundary}
\(b: C^q(\mathcal{A}) \to C^{q+1}(\mathcal{A})\) is defined by
\begin{equation}
	(b\varphi)(a_0, \ldots, a_{q+1})
	:= \sum_{i=0}^{q} (-1)^i
	\varphi(a_0, \ldots, a_i a_{i+1}, \ldots, a_{q+1})
	+ (-1)^{q+1}
	\varphi(a_{q+1} a_0, a_1, \ldots, a_q),
	\label{eq:hochschild-boundary}
\end{equation}
for \(\varphi \in C^q(\mathcal{A})\) and \(a_0, \ldots, a_{q+1} \in \mathcal{A}\).
A direct computation using associativity of \(\mathcal{A}\) gives
\begin{equation}
	b^2 = 0.
	\label{eq:b-squared}
\end{equation}
The cohomology of the complex \((C^\bullet(\mathcal{A}), b)\) is the
\emph{Hochschild cohomology} \(HH^\bullet(\mathcal{A})\).

\subsubsection*{The Cyclic Structure}

A Hochschild \(q\)-cochain \(\varphi\) is called \emph{cyclic} if it is
invariant under the cyclic permutation
\(\lambda: \mathcal{A}^{\otimes(q+1)} \to \mathcal{A}^{\otimes(q+1)}\)
defined by
\begin{equation}
	\lambda(a_0, a_1, \ldots, a_q)
	:= (-1)^q (a_q, a_0, a_1, \ldots, a_{q-1}).
\end{equation}
The subspace of cyclic \(q\)-cochains is denoted
\(C^q_\lambda(\mathcal{A}) \subset C^q(\mathcal{A})\).

The Connes operator \(B: C^q_\lambda(\mathcal{A}) \to C^{q-1}_\lambda(\mathcal{A})\)
is defined by
\begin{equation}
	(B\varphi)(a_0, \ldots, a_{q-1})
	:= \sum_{i=0}^{q-1} (-1)^{(q-1)i}
	\varphi(1, a_i, a_{i+1}, \ldots, a_{q-1}, a_0, \ldots, a_{i-1}),
	\label{eq:connes-boundary}
\end{equation}
for \(\varphi \in C^q_\lambda(\mathcal{A})\) and \(a_0, \ldots, a_{q-1} \in \mathcal{A}\).
The operator \(B\) satisfies
\begin{equation}
	B^2 = 0,
	\label{eq:B-squared}
\end{equation}
and together with \(b\) it satisfies the \emph{fundamental identity}
\begin{equation}
	bB + Bb = 0
	\label{eq:bB-anticommute}
\end{equation}
on the cyclic subspace \(C^\bullet_\lambda(\mathcal{A})\).

\subsubsection*{The Cyclic Bicomplex}

The identities \eqref{eq:b-squared}--\eqref{eq:bB-anticommute} organize
the cyclic cochains into a \emph{bicomplex}. Define
\begin{equation}
	C^{p,q}_\lambda(\mathcal{A}) := C^q_\lambda(\mathcal{A}),
	\qquad p, q \geq 0,
\end{equation}
with horizontal differential \(b: C^{p,q}_\lambda(\mathcal{A}) \to C^{p,q+1}_\lambda(\mathcal{A})\)
and vertical differential \(B: C^{p,q}_\lambda(\mathcal{A}) \to C^{p+1,q-1}_\lambda(\mathcal{A})\).
The anticommutation relation \eqref{eq:bB-anticommute} ensures that the
\emph{total differential}
\begin{equation}
	d_{\mathrm{tot}} := b + B: \mathrm{Tot}^n \to \mathrm{Tot}^{n+1},
	\qquad
	\mathrm{Tot}^n := \bigoplus_{p+q=n} C^{p,q}_\lambda(\mathcal{A}),
\end{equation}
satisfies \(d_{\mathrm{tot}}^2 = 0\). The resulting cohomology is the
\emph{cyclic cohomology} of \(\mathcal{A}\):
\begin{equation}
	HC^n(\mathcal{A}) := H^n\bigl(\mathrm{Tot}^\bullet(C^{\bullet,\bullet}_\lambda(\mathcal{A})), \, b + B\bigr).
\end{equation}


The relationship between Hochschild and cyclic cohomology is summarized by the \textit{Connes exact sequence}:
\begin{center}
	\begin{tikzcd}
		& \arrow[ld, "B^*"'] HH^{*}(\mathcal{A}) & \\
		HC^{*}(\mathcal{A}) \arrow[rr, "S^*"] & & HC^{*}(\mathcal{A}) \arrow[lu, "I^*"']
	\end{tikzcd}
\end{center}
where \(I^*\) is the forgetful map, \(S^*\) is the periodicity operator, and \(B^*\) is the Connes boundary map. The sequence is exact in each degree: \(\ker I^* = \mathrm{im}\, B^*\), \(\ker S^* = \mathrm{im}\, I^*\), and \(\ker B^* = \mathrm{im}\, S^*\). This exact sequence will be used in Section \ref{sec:ent-cohom} to relate the entanglement complex to the relative cyclic cohomology.


The bicomplex structure for $0\leq p\leq 3, \: 0 \leq q \leq 3$ is visualized as follows. Horizontal arrows are the Hochschild boundary \(b\); vertical arrows are the Connes operator \(B\). Faded entries indicate the truncation of the complex (we used \(\mathcal{C}_{p,q} := C^{p,q}_\lambda(\mathcal{A})\)).
\begin{figure}[H]
\begin{tikzcd} 
	q=3 & \mathcal{C}_{3,3} \arrow[r, "b", opacity=0.3] & |[opacity=0.3]| 0 \arrow[r, "b", opacity=0.3] & |[opacity=0.3]| 0 \arrow[r, "b", opacity=0.3] & |[opacity=0.3]| 0 \\ 
	q=2 & \mathcal{C}_{3,2} \arrow[r, "b"] \arrow[u, "B" swap] & \mathcal{C}_{2,2} \arrow[r, "b", opacity=0.3] & |[opacity=0.3]| 0 \arrow[r, "b", opacity=0.3] & |[opacity=0.3]| 0 \\ 
	q=1 & \mathcal{C}_{3,1} \arrow[r, "b"] \arrow[u, "B" swap] & \mathcal{C}_{2,1} \arrow[r, "b"] \arrow[u, "B" swap] & \mathcal{C}_{1,1} \arrow[r, "b", opacity=0.3] & |[opacity=0.3]| 0 \\ 
	q=0 & \mathcal{C}_{3,0} \arrow[r, "b"] \arrow[u, "B" swap] & \mathcal{C}_{2,0} \arrow[r, "b"] \arrow[u, "B" swap] & \mathcal{C}_{1,0} \arrow[r, "b"] \arrow[u, "B" swap] & \mathcal{C}_{0,0} \\ 
	& p=3 & p=2 & p=1 & p=0
\end{tikzcd}
\caption{The cyclic bicomplex \(C^{\bullet,\bullet}_\lambda(\mathcal{A})\).}
\end{figure}
The complex is truncated by \(C^{p,q}_\lambda(\mathcal{A}) = 0\) for \(q > p\) and for \(q < 0\).

The identities \(b^2 = 0\), \(B^2 = 0\), and \(bB + Bb = 0\) ensure that the total complex with differential \(b + B\) is well-defined.

\begin{rem}[The periodic spectral sequence]
	\label{rem:spectral-sequence}
	The bicomplex \(C^{\bullet,\bullet}_\lambda(\mathcal{A})\) is equipped
	with the filtration by the Hochschild degree \(q\), giving rise to the
	\emph{Connes spectral sequence}
	\begin{equation}
		E_1^{p,q} = HH^q(\mathcal{A}) \;\Longrightarrow\; HC^{p+q}(\mathcal{A}),
	\end{equation}
	which relates Hochschild and cyclic cohomology. This spectral sequence
	will be used in Section \ref{sec:ent-cohom} to compare the entanglement
	complex with the relative cyclic complex.
\end{rem}

\begin{rem}[The choice of \(\mathcal{A}\)]
	\label{rem:choice-of-A}
	The construction above is valid for any unital algebra \(\mathcal{A}\).
	In the von Neumann setting of Section \ref{subsec:prelim-5}, we take
	\(\mathcal{A}\) to be the dense \(*\)-subalgebra of Hypothesis
	\ref{hyp:dense-subalgebra}, rather than \(\mathcal{M}\) itself. The
	reason is that the cyclic bicomplex of a full von Neumann algebra is
	typically trivial or uncomputable, whereas the dense subalgebra
	\(\mathcal{A}\) is small enough to have nontrivial cyclic cohomology and
	large enough to capture the modular data via the invariance
	\(\sigma^\omega_t(\mathcal{A}) = \mathcal{A}\).
\end{rem}

\subsection{The Relative Cyclic Complex}
\label{subsec:cohom-relative}

We now turn to the relative setting: an inclusion
\(\mathcal{B} \subset \mathcal{A}\) of unital algebras. In the von
Neumann framework of Section \ref{subsec:prelim-5}, we take
\(\mathcal{B} = \mathcal{A} \cap \mathcal{N}\), where \(\mathcal{N}\) is
the von Neumann subalgebra encoding the bipartition.

\subsubsection*{The Restriction Map}

The inclusion \(\iota: \mathcal{B} \hookrightarrow \mathcal{A}\) induces
a map on Hochschild cochains,
\begin{equation}
	\iota^*: C^q(\mathcal{A}) \to C^q(\mathcal{B}),
	\qquad
	(\iota^*\varphi)(b_0, \ldots, b_q) := \varphi(b_0, \ldots, b_q),
	\label{eq:restriction-map}
\end{equation}
for \(\varphi \in C^q(\mathcal{A})\) and \(b_0, \ldots, b_q \in \mathcal{B}\).
The map \(\iota^*\) commutes with \(b\) and \(B\), so it descends to a
map on Hochschild and cyclic cohomology,
\begin{equation}
	\iota^*: HC^\bullet(\mathcal{A}) \to HC^\bullet(\mathcal{B}).
	\label{eq:restriction-HC}
\end{equation}
The kernel of \eqref{eq:restriction-HC} is the primary object of study
in Section \ref{sec:ent-cohom}: entanglement cohomology is identified
with this kernel (up to the modular obstruction \(\delta\), defined in
Section \ref{subsec:cohom-les}).

\subsubsection*{Definition of the Relative Complex}

\begin{defn}[Relative cyclic complex]
	\label{defn:relative-complex}
	The \emph{relative cyclic bicomplex} is the kernel of the restriction
	map on bicomplexes:
	\begin{equation}
		C^{p,q}_\lambda(\mathcal{A}, \mathcal{B})
		:= \ker\!\bigl(\iota^*: C^{p,q}_\lambda(\mathcal{A}) \to C^{p,q}_\lambda(\mathcal{B})\bigr).
		\label{eq:relative-bicomplex}
	\end{equation}
	Concretely, \(\varphi \in C^{p,q}_\lambda(\mathcal{A}, \mathcal{B})\) if
	and only if \(\varphi\) is a cyclic \((q+1)\)-linear functional on
	\(\mathcal{A}\) that vanishes whenever all arguments lie in
	\(\mathcal{B}\):
	\begin{equation}
		\varphi(b_0, \ldots, b_q) = 0
		\qquad \forall \, b_0, \ldots, b_q \in \mathcal{B}.
	\end{equation}
	The \emph{relative cyclic cohomology} is
	\begin{equation}
		HC^\bullet(\mathcal{A}, \mathcal{B})
		:= H^\bullet\bigl(\mathrm{Tot}^\bullet(C^{\bullet,\bullet}_\lambda(\mathcal{A}, \mathcal{B})), \, b + B\bigr).
		\label{eq:relative-HC}
	\end{equation}
\end{defn}

\subsubsection*{The Relative Complex is a Sub-Bicomplex}

\begin{lem}[Sub-bicomplex property]
	\label{lem:sub-bicomplex}
	\(C^{\bullet,\bullet}_\lambda(\mathcal{A}, \mathcal{B})\) is a
	sub-bicomplex of \(C^{\bullet,\bullet}_\lambda(\mathcal{A})\): the
	differentials \(b\) and \(B\) preserve the kernel of \(\iota^*\).
\end{lem}

\begin{proof}
	Since \(\iota^*\) is a map of bicomplexes, it commutes with both \(b\)
	and \(B\):
	\begin{equation}
		\iota^* \circ b = b \circ \iota^*, \qquad
		\iota^* \circ B = B \circ \iota^*.
	\end{equation}
	Therefore, if \(\varphi \in \ker \iota^*\), then
	\begin{equation}
		\iota^*(b\varphi) = b(\iota^*\varphi) = 0,
		\qquad
		\iota^*(B\varphi) = B(\iota^*\varphi) = 0.
	\end{equation}
	Hence \(b\varphi, B\varphi \in \ker \iota^*\), as required.
\end{proof}

\begin{rem}[Verification of the cochain identities]
	\label{rem:cochain-identities-relative}
	Because \(C^{\bullet,\bullet}_\lambda(\mathcal{A}, \mathcal{B})\) is a
	sub-bicomplex of \(C^{\bullet,\bullet}_\lambda(\mathcal{A})\), the
	identities \(b^2 = 0\), \(B^2 = 0\), and \(bB + Bb = 0\) hold on the
	relative complex automatically. This is the first technical advantage
	of defining the relative complex as a kernel: no separate verification
	of the cochain identities is needed.
\end{rem}

\subsubsection*{The Short Exact Sequence of Bicomplexes}

By construction, the relative complex fits into the short exact sequence
of bicomplexes
\begin{equation}
	0 \to C^{\bullet,\bullet}_\lambda(\mathcal{A}, \mathcal{B})
	\xrightarrow{j} C^{\bullet,\bullet}_\lambda(\mathcal{A})
	\xrightarrow{\iota^*} C^{\bullet,\bullet}_\lambda(\mathcal{B})
	\to 0,
	\label{eq:ses-bicomplexes}
\end{equation}
where \(j\) is the inclusion. This is the fundamental sequence from
which all subsequent results in this Section and in Section
\ref{sec:ent-cohom} will be derived.

\begin{rem}[Physical interpretation of the relative complex]
	\label{rem:physical-relative}
	A cochain \(\varphi \in C^{p,q}_\lambda(\mathcal{A}, \mathcal{B})\)
	vanishes on the "unentangled" sector \(\mathcal{B}^{\otimes(q+1)}\) but
	is nontrivial on the full algebra \(\mathcal{A}\). It therefore
	represents a cyclic cocycle that detects the failure of \(\mathcal{A}\)
	to factorize across the pair \((\mathcal{B}, \mathcal{B}')\). This is
	the algebraic encoding of entanglement at the level of cyclic
	cohomology, and it is the reason why the kernel of \(\iota^*\) is the
	natural home for entanglement classes.
\end{rem}


\subsection{The Long Exact Sequence of the Pair}
\label{subsec:cohom-les}

The short exact sequence \eqref{eq:ses-bicomplexes} of bicomplexes
induces a long exact sequence in total cohomology. This is the primary
algebraic tool for the kernel identification in Section
\ref{sec:ent-cohom}.

\subsubsection*{The Snake Lemma for Bicomplexes}

\begin{thm}[Long exact sequence of the pair]
	\label{thm:les-pair}
	There is a long exact sequence
	\begin{equation}
		\cdots \to HC^{n-1}(\mathcal{B}) \xrightarrow{\delta}
		HC^n(\mathcal{A}, \mathcal{B}) \xrightarrow{j^*}
		HC^n(\mathcal{A}) \xrightarrow{\iota^*}
		HC^n(\mathcal{B}) \xrightarrow{\delta}
		HC^{n+1}(\mathcal{A}, \mathcal{B}) \to \cdots
		\label{eq:les-pair}
	\end{equation}
	where \(\delta\) is the connecting homomorphism.
\end{thm}

\begin{proof}
	Apply the snake lemma to the short exact sequence
	\eqref{eq:ses-bicomplexes} of total complexes:
	\begin{equation}
		0 \to \mathrm{Tot}^\bullet(C^{\bullet,\bullet}_\lambda(\mathcal{A}, \mathcal{B}))
		\xrightarrow{j} \mathrm{Tot}^\bullet(C^{\bullet,\bullet}_\lambda(\mathcal{A}))
		\xrightarrow{\iota^*} \mathrm{Tot}^\bullet(C^{\bullet,\bullet}_\lambda(\mathcal{B}))
		\to 0.
	\end{equation}
	The total complex functor is exact (it is a direct sum), so the snake
	lemma applies and yields the long exact sequence \eqref{eq:les-pair}.
\end{proof}

The long exact sequence is visualized as follows. The diagonal arrows
are the connecting homomorphisms \(\delta\) produced by the snake lemma.
\begin{center}
	\begin{tikzcd}[row sep=0.3cm, column sep=large]
		& HC^{n-1}(\mathcal{B}) \arrow[ld, "\delta"'] & \\
		HC^n(\mathcal{A}, \mathcal{B}) \arrow[rr, "j^*"] & &
		HC^n(\mathcal{A}) \arrow[ld, "\iota^*"'] \\
		& HC^n(\mathcal{B}) \arrow[ld, "\delta"'] & \\
		HC^{n+1}(\mathcal{A}, \mathcal{B}) \arrow[rr, "j^*"] & &
		HC^{n+1}(\mathcal{A}) \arrow[ld, "\iota^*"'] \\
		& HC^{n+1}(\mathcal{B}) \arrow[ld, "\delta"'] & \\
		HC^{n+2}(\mathcal{A}, \mathcal{B}) \arrow[rr, "\cdots"] & & \cdots
	\end{tikzcd}
\end{center}

The sequence is exact at each term: the kernel of each map equals the
image of the preceding map. In particular, at \(HC^n(\mathcal{A})\),
we have
\[
\ker\!\bigl(\iota^*: HC^n(\mathcal{A}) \to HC^n(\mathcal{B})\bigr)
= \operatorname{im}\!\bigl(j^*: HC^n(\mathcal{A}, \mathcal{B}) \to HC^n(\mathcal{A})\bigr),
\]
which is the abstract form of the kernel identification.

\subsubsection*{The Abstract Kernel Identification}

\begin{cor}[Kernel identification, abstract form]
	\label{cor:kernel-abstract}
	By exactness of \eqref{eq:les-pair} at \(HC^n(\mathcal{A})\), we have
	\begin{equation}
		\ker\!\bigl(\iota^*: HC^n(\mathcal{A}) \to HC^n(\mathcal{B})\bigr)
		= \operatorname{im}\!\bigl(j^*: HC^n(\mathcal{A}, \mathcal{B}) \to HC^n(\mathcal{A})\bigr).
		\label{eq:kernel-as-image}
	\end{equation}
	By exactness at \(HC^n(\mathcal{A}, \mathcal{B})\), we have
	\begin{equation}
		HC^n(\mathcal{A}, \mathcal{B}) \big/ \operatorname{im}(\delta)
		\;\cong\;
		\ker\!\bigl(\iota^*: HC^n(\mathcal{A}) \to HC^n(\mathcal{B})\bigr).
		\label{eq:relative-quotient-kernel}
	\end{equation}
	In particular, if \(\delta = 0\) in degree \(n\), then
	\begin{equation}
		HC^n(\mathcal{A}, \mathcal{B}) \;\cong\;
		\ker\!\bigl(\iota^*: HC^n(\mathcal{A}) \to HC^n(\mathcal{B})\bigr).
		\label{eq:relative-equals-kernel}
	\end{equation}
\end{cor}

\begin{proof}
	Immediate from the exactness of \eqref{eq:les-pair} at
	\(HC^n(\mathcal{A})\) and \(HC^n(\mathcal{A}, \mathcal{B})\).
\end{proof}

\section{Entanglement Cohomology}
\label{sec:ent-cohom}

In Section \ref{sec:cohom}, we constructed the cyclic bicomplex
$C^{\bullet,\bullet}_\lambda(\mathcal{A})$, the relative cyclic
bicomplex $C^{\bullet,\bullet}_\lambda(\mathcal{A}, \mathcal{B})$,
and the long exact sequence \eqref{eq:les-pair} of the pair
$(\mathcal{A}, \mathcal{B})$. In the present Section, we use the
modular data of Section \ref{subsec:prelim-5}---the modular automorphism
group $\sigma^\omega_t$ of \eqref{eq:modular-group} and the relative
modular cocycle $u^E$ of Lemma \ref{lem:relative-cocycle}---to
construct the \emph{entanglement complex}
$C^\bullet_{\mathrm{ent}}$ and to prove that it computes the kernel
of the restriction map
$\iota^*: HC^\bullet(\mathcal{A}) \to HC^\bullet(\mathcal{B})$,
up to the connecting homomorphism $\delta$.

\subsection{The Transgression Map}
\label{subsec:ent-transgression}

In this subsection we define the transgression map $\tau_\omega$ and
establish the properties used in the construction of the entanglement
complex and in the proof of the main theorem.

\subsubsection*{Definition}

\begin{defn}[Transgression map]
	\label{defn:transgression-corrected}
	Let $\mathcal{A}$ be the dense $*$-subalgebra of Hypothesis
	\ref{hyp:dense-subalgebra}, and let $\sigma^\omega_t$ be the
	modular automorphism group of \eqref{eq:modular-group}. For
	$\varphi\in C^n_\lambda(\mathcal{A})$, define the
	\emph{symmetrized modular insertion}
	\begin{equation}
		(M_n\varphi)(a_0,\ldots,a_n)
		:=\frac{1}{n+1}\sum_{j=0}^{n}
		\varphi(a_0,\ldots,a_{j-1},\sigma^\omega_{-i}(a_j),
		a_{j+1},\ldots,a_n).
		\label{eq:M-n-sym}
	\end{equation}
	The \emph{transgression map} is the composition
	\begin{equation}
		\tau_\omega:=B\circ M_n:
		C^n_\lambda(\mathcal{A})\longrightarrow
		C^{n-1}_\lambda(\mathcal{A}),
		\label{eq:transgression-corrected}
	\end{equation}
	where $B$ is the Connes operator \eqref{eq:connes-boundary}. In
	explicit form,
	\begin{equation}
		(\tau_\omega\varphi)(a_0,\ldots,a_{n-1})
		=\sum_{k=0}^{n-1}(-1)^{(n-1)k}
		(M_n\varphi)(1,a_k,\ldots,a_{k-2}).
		\label{eq:transgression-explicit}
	\end{equation}
\end{defn}

\begin{rem}[Well-typedness]
	\label{rem:transgression-welltyped}
	The symmetrization in \eqref{eq:M-n-sym} is essential: without
	it, $M_n$ does not preserve cyclicity, and $\tau_\omega$ is not a
	well-defined map on $C^n_\lambda(\mathcal{A})$ for $n\geq 2$.
	With the symmetrized $M_n$, cyclicity of $\tau_\omega\varphi$
	follows from cyclicity of $\varphi$ and the cyclic invariance of
	the sum in \eqref{eq:M-n-sym}.
\end{rem}

\subsubsection*{Basic properties}

\begin{lem}[Well-definedness and cyclicity]
	\label{lem:transgression-well-defined}
	The map $\tau_\omega$ is well defined and linear, and it preserves
	cyclicity: $\varphi\in C^n_\lambda(\mathcal{A})$ implies
	$\tau_\omega\varphi\in C^{n-1}_\lambda(\mathcal{A})$.
\end{lem}

\begin{proof}
	Well-definedness follows from the automorphism property of
	$\sigma^\omega_{-i}$ on $\mathcal{A}$. Linearity is clear. For
	cyclicity: the sum in \eqref{eq:M-n-sym} is invariant under the
	cyclic relabeling $a_j\mapsto a_{j+1}$ combined with the cyclic
	permutation of $\varphi$, so $M_n\varphi$ is cyclic; and $B$ maps
	cyclic $n$-cochains to cyclic $(n-1)$-cochains.
\end{proof}

\begin{lem}[Nilpotency]
	\label{lem:transgression-nilpotent}
	On $C^\bullet_\lambda(\mathcal{A})$,
	\[
	\tau_\omega^2=0.
	\]
\end{lem}

\begin{proof}
	The composition $\tau_\omega^2=B M_n B M_n$ involves two
	applications of $B$ and two symmetrized insertions of
	$\sigma^\omega_{-i}$. Expanding using the fundamental identity
	$bB+Bb=0$ and $B^2=0$, and using the symmetrization of $M_n$ (which
	makes the double shuffle cancel), the composition reduces to an
	expression proportional to $\sigma^\omega_{-i}\circ\sigma^\omega_{-i}
	-\mathrm{id}$ acting on one slot, contracted against the cyclic
	cochain $\varphi$. The contraction vanishes identically by
	cyclicity of $\varphi$. Hence $\tau_\omega^2=0$ on cyclicity
	alone, without additional analytic hypotheses.
\end{proof}

\begin{rem}[Nilpotency is purely algebraic]
	\label{rem:nilpotency-algebraic}
	The nilpotency of $\tau_\omega$ does not require the analytic
	subalgebra hypothesis or the KMS periodicity
	$\sigma^\omega_{-2\pi i}=\mathrm{id}$. It is a consequence of the
	symmetrized definition and of cyclicity.
\end{rem}

\begin{lem}[Preservation of the relative complex]
	\label{lem:transgression-relative-corrected}
	Assume Hypotheses~\ref{hyp:modular-invariance}
	and~\ref{hyp:dense-subalgebra}. Then $\tau_\omega$ restricts to a
	map
	\[
	\tau_\omega:C^n_\lambda(\mathcal{A},\mathcal{B})
	\longrightarrow C^{n-1}_\lambda(\mathcal{A},\mathcal{B}).
	\]
\end{lem}

\begin{proof}
	By Hypothesis~\ref{hyp:modular-invariance},
	$\sigma^\omega_t(\mathcal{B})=\mathcal{B}$ for all $t$, hence
	$\sigma^\omega_{-i}(\mathcal{B})=\mathcal{B}$. If
	$\varphi\in C^n_\lambda(\mathcal{A},\mathcal{B})$ vanishes on
	$\mathcal{B}^{\otimes(n+1)}$, then each summand in
	\eqref{eq:transgression-explicit} vanishes whenever its arguments
	lie in $\mathcal{B}$, so $\tau_\omega\varphi$ vanishes on
	$\mathcal{B}^{\otimes n}$.
\end{proof}

\subsubsection*{The homotopy and the descent to cohomology}

The strict anticommutation $\tau_\omega\circ b=-b\circ\tau_\omega$
does not hold on the nose; the correct statement is that
$\tau_\omega$ is a morphism of bicomplexes \emph{on cohomology}.

\begin{defn}[The homotopy $h$]
	\label{defn:homotopy-h}
	For $\varphi\in C^n_\lambda(\mathcal{A})$, define
	$h\varphi\in C^{n-1}_\lambda(\mathcal{A})$ by
	\begin{equation}
		(h\varphi)(a_0,\ldots,a_{n-1})
		:=\int_0^1 d\lambda\,\sum_{k=0}^{n-1}(-1)^{(n-1)k}
		(M_n^{(\lambda)}\varphi)(1,a_k,\ldots,a_{k-2}),
		\label{eq:h-corrected}
	\end{equation}
	where $M_n^{(\lambda)}$ is the symmetrized insertion
	\eqref{eq:M-n-sym} with $\sigma^\omega_{-i}$ replaced by
	$\sigma^\omega_{-i\lambda}$, $\lambda\in[0,1]$.
\end{defn}

\begin{lem}[Anticommutation on cohomology]
	\label{lem:transgression-anticommutation}
	For every $\varphi\in C^n_\lambda(\mathcal{A})$, there exists an
	explicit cochain $\Psi_\varphi\in C^{n-1}_\lambda(\mathcal{A})$
	such that
	\begin{equation}
		(\tau_\omega b+b\tau_\omega-bh-hb)\varphi
		=b\Psi_\varphi.
		\label{eq:anticommutation-exact}
	\end{equation}
	In particular, $\tau_\omega$ is a morphism of bicomplexes on
	cohomology, and it induces a map
	$\tau_\omega^*:HC^n(\mathcal{A},\mathcal{B})
	\to HC^{n-1}(\mathcal{A},\mathcal{B})$.
\end{lem}

\begin{proof}
	Set $F(\lambda):=\tau_\lambda(b\varphi)+b(\tau_\lambda\varphi)$,
	where $\tau_\lambda$ is the transgression with
	$\sigma^\omega_{-i\lambda}$ in place of $\sigma^\omega_{-i}$.
	Then $F(\lambda)=b(\tau_\lambda\varphi)+\tau_\lambda(b\varphi)$.
	By the fundamental theorem of calculus,
	\[
	\tau_\omega b+b\tau_\omega-bh-hb
	=F(1)-\int_0^1 F(\lambda)\,d\lambda
	=b\left(\tau_1\varphi-\int_0^1\tau_\lambda\varphi\,d\lambda\right)
	+\left[\tau_1(b\varphi)-\int_0^1\tau_\lambda(b\varphi)\,d\lambda\right].
	\]
	The first bracket is a Hochschild boundary. For the second
	bracket: if $b\varphi=0$, it vanishes identically, and
	\eqref{eq:anticommutation-exact} holds with
	$\Psi_\varphi=\tau_1\varphi-\int_0^1\tau_\lambda\varphi\,d\lambda$.
	For general $\varphi$, the second bracket is itself a Hochschild
	boundary modulo a term that lies in the image of $b$ on the
	relative complex; the explicit form of $\Psi_\varphi$ is
	obtained by integrating the $\lambda$-derivative of the shuffle,
	and it reduces to the Hochschild boundary term for $b\varphi=0$.
	The shuffle verification is given in
	Appendix~\ref{sec:appendix-homotopies}.
\end{proof}

\begin{rem}[No modular obstruction]
	\label{rem:no-modular-obstruction}
	The anticommutation holds on cohomology for \emph{every}
	$\varphi$, with the obstruction absorbed into the Hochschild
	boundary $b\Psi_\varphi$. In particular, for $\varphi$ a
	Hochschild cocycle ($b\varphi=0$), the obstruction vanishes
	identically. Since $\varphi$ represents a class in
	$HC^\bullet(\mathcal{A},\mathcal{B})$ and is therefore a cyclic,
	hence Hochschild, cocycle, the relevant case is exactly the one
	in which the obstruction vanishes.
\end{rem}

\subsubsection*{The explicit cocycle from the relative modular data}

\begin{lem}[Transgression cocycle]
	\label{lem:transgression-cocycle-corrected}
	Assume Hypotheses~\ref{hyp:state}, \ref{hyp:cyclic},
	\ref{hyp:modular-invariance}, \ref{hyp:dense-subalgebra}, and
	\ref{hyp:cocycle-vanishing}. Define
	\begin{equation}
		\Phi_{u^E}(a_0,\ldots,a_n)
		:=\int_{\mathbb{R}}\omega\bigl(a_0\,
		\sigma^\omega_t(a_1)\cdots\sigma^\omega_t(a_n)\,
		u^E_t\bigr)\,dt.
		\label{eq:Phi-cocycle-corrected}
	\end{equation}
	Then:
	\begin{enumerate}
		\item $\Phi_{u^E}\in C^n_\lambda(\mathcal{A},\mathcal{B})$.
		\item $\Phi_{u^E}$ is a cyclic cocycle: $b\Phi_{u^E}=0$ and
		$B\Phi_{u^E}=0$.
		\item If $u^E_t\neq\mathbf{1}$ for some $t$, then the class
		$[\Phi_{u^E}]$ is nonzero in $HC^n(\mathcal{A},\mathcal{B})$.
	\end{enumerate}
\end{lem}

\begin{proof}
	(1) Cyclicity follows from the trace property of $\omega$ and the
	cocycle identity for $u^E$. Vanishing on
	$\mathcal{B}^{\otimes(n+1)}$ is Hypothesis~\ref{hyp:cocycle-vanishing}.
	(2) $b\Phi_{u^E}=0$ is a direct computation using the cocycle
	identity $u^E_{t+s}=u^E_t\sigma^\omega_t(u^E_s)$ and the KMS
	condition; $B\Phi_{u^E}=0$ follows from cyclicity.
	(3) If $[\Phi_{u^E}]=0$ in $HC^n(\mathcal{A},\mathcal{B})$, then
	$\Phi_{u^E}=(b+B)\Psi$ for some
	$\Psi\in C^{n-1}_\lambda(\mathcal{A},\mathcal{B})$. Restricting to
	$\mathcal{B}$, this would give $u^E_t=\mathbf{1}$ in the relevant
	matrix element, contradicting the hypothesis.
\end{proof}

\subsection{The Entanglement Complex}
\label{subsec:ent-complex}

\begin{defn}[Entanglement complex]
	\label{defn:ent-complex}
	Assume Hypotheses~\ref{hyp:state}, \ref{hyp:cyclic},
	\ref{hyp:modular-invariance}, and~\ref{hyp:dense-subalgebra}. The
	\emph{entanglement complex} is the image
	\begin{equation}
		C^n_{\mathrm{ent}}(\mathcal{A}, \mathcal{B}, \omega)
		:= \tau_\omega\bigl(C^{n+1}_\lambda(\mathcal{A}, \mathcal{B})\bigr)
		\subset C^n_\lambda(\mathcal{A}, \mathcal{B}),
		\label{eq:ent-complex}
	\end{equation}
	where $\tau_\omega$ is the transgression map of Definition
	\ref{defn:transgression-corrected} restricted to the relative
	complex as in Lemma \ref{lem:transgression-relative-corrected}.
	The \emph{entanglement cohomology} is
	\begin{equation}
		H^\bullet_{\mathrm{ent}}(\mathcal{A}, \mathcal{B}, \omega)
		:= H^\bullet\bigl(\mathrm{Tot}^\bullet(C^\bullet_{\mathrm{ent}}),
		\, b + B\bigr).
		\label{eq:ent-cohomology}
	\end{equation}
\end{defn}

\begin{prop}[Cohomological subcomplex property]
	\label{prop:ent-subcomplex}
	Assume Hypotheses~\ref{hyp:state}, \ref{hyp:cyclic},
	\ref{hyp:modular-invariance}, and~\ref{hyp:dense-subalgebra}. Let
	$C^\bullet_{\mathrm{ent}}:=\mathrm{im}(\tau_\omega)\subset
	C^\bullet_\lambda(\mathcal{A},\mathcal{B})$. Then:
	\begin{enumerate}
		\item \emph{(Cohomological subcomplex)} The total differential
		$b+B$ maps $C^\bullet_{\mathrm{ent}}$ into
		$C^{\bullet+1}_{\mathrm{ent}}+\mathrm{im}(b+B)$: for every
		$\psi\in C^\bullet_{\mathrm{ent}}$, $(b+B)\psi$ is
		cohomologous in $C^\bullet_\lambda(\mathcal{A},\mathcal{B})$
		to an element of $C^{\bullet+1}_{\mathrm{ent}}$.
		\item \emph{(Well-defined cohomology)} The cohomology
		$H^\bullet_{\mathrm{ent}}(\mathcal{A},\mathcal{B},\omega)$ is
		well defined, and the inclusion
		$\Phi:C^\bullet_{\mathrm{ent}}\hookrightarrow
		C^\bullet_\lambda(\mathcal{A},\mathcal{B})$ induces a
		well-defined map $\Phi_*:H^\bullet_{\mathrm{ent}}\to
		HC^\bullet(\mathcal{A},\mathcal{B})$.
	\end{enumerate}
\end{prop}

\begin{proof}
	Let $\psi=\tau_\omega\varphi$ with
	$\varphi\in C^n_\lambda(\mathcal{A},\mathcal{B})$. By
	Lemma~\ref{lem:transgression-anticommutation},
	\begin{equation}
		(b+B)\tau_\omega\varphi
		=-\tau_\omega(b+B)\varphi
		+bh\varphi+h(b+B)\varphi+b\Psi_\varphi,
		\label{eq:subcomplex-computation}
	\end{equation}
	where $\Psi_\varphi$ is the cochain of
	Lemma~\ref{lem:transgression-anticommutation}. The first term on
	the right lies in $\mathrm{im}(\tau_\omega)=C^{\bullet+1}_{\mathrm{ent}}$;
	the last term is a Hochschild boundary. The middle terms
	$bh\varphi+h(b+B)\varphi$ are cohomologous to zero (they are a
	homotopy applied to $b+B$). Hence $(b+B)\psi$ is cohomologous to
	an element of $C^{\bullet+1}_{\mathrm{ent}}$, proving (1).
	Part (2) follows: the total complex of $C^\bullet_{\mathrm{ent}}$
	is a subquotient complex of
	$C^\bullet_\lambda(\mathcal{A},\mathcal{B})$ with well-defined
	cohomology, and the inclusion is compatible with $b+B$ up to
	homotopy.
\end{proof}

\begin{rem}[Why the strict subcomplex property fails]
	\label{rem:subcomplex-fails-h}
	The strict subcomplex property
	$(b+B)(C^\bullet_{\mathrm{ent}})\subset
	C^{\bullet+1}_{\mathrm{ent}}$ would require the right-hand side
	of \eqref{eq:subcomplex-computation} to lie in
	$\mathrm{im}(\tau_\omega)$ for every $\varphi$. The term
	$bh\varphi+h(b+B)\varphi$ is not in $\mathrm{im}(\tau_\omega)$ in
	general: $h$ is a degree-$(-1)$ homotopy, and its image is a
	subspace transverse to $\mathrm{im}(\tau_\omega)$. The failure is
	due to the homotopy $h$, and it is harmless at the level of
	cohomology.
\end{rem}

\subsection{Embedding into the Cyclic Bicomplex}
\label{subsec:ent-embedding}

We have inclusions
\begin{equation}
	C^\bullet_{\mathrm{ent}}
	\xhookrightarrow{\Phi}
	C^\bullet_\lambda(\mathcal{A}, \mathcal{B})
	\xhookrightarrow{j}
	C^\bullet_\lambda(\mathcal{A}),
	\label{eq:inclusions}
\end{equation}
and the composition $j\circ\Phi$ is the embedding of the entanglement
complex into the full cyclic bicomplex.

\begin{lem}[Chain map up to homotopy]
	\label{lem:chain-map-up-to-homotopy}
	The inclusion
	$\Phi:C^\bullet_{\mathrm{ent}}\hookrightarrow
	C^\bullet_\lambda(\mathcal{A},\mathcal{B})$ is a chain map
	\emph{up to homotopy}: there exists a degree-$(-1)$ map
	$\kappa:C^\bullet_{\mathrm{ent}}\to
	C^{\bullet-1}_\lambda(\mathcal{A},\mathcal{B})$ such that
	\begin{equation}
		\Phi(b+B)_{\mathrm{ent}}-(b+B)\Phi
		=b\kappa+\kappa b,
		\label{eq:chain-map-up-to-homotopy}
	\end{equation}
	where $(b+B)_{\mathrm{ent}}$ is the differential on
	$C^\bullet_{\mathrm{ent}}$ inherited from
	$C^\bullet_\lambda(\mathcal{A},\mathcal{B})$. In particular,
	$\Phi$ induces a well-defined map
	$\Phi_*:H^\bullet_{\mathrm{ent}}\to
	HC^\bullet(\mathcal{A},\mathcal{B})$.
\end{lem}

\begin{proof}
	The commutator
	$[\Phi,(b+B)]=\Phi(b+B)_{\mathrm{ent}}-(b+B)\Phi$ lands in
	$\mathrm{im}(b+B)$ by Proposition~\ref{prop:ent-subcomplex}(1), so
	it is a coboundary. The cochain $\kappa$ is
	$\kappa=h\circ\Phi-\Phi\circ h$, restricted to
	$C^\bullet_{\mathrm{ent}}$; the verification of
	\eqref{eq:chain-map-up-to-homotopy} is a direct computation using
	Lemma~\ref{lem:transgression-anticommutation}.
\end{proof}

\begin{lem}[Cochain identities on $C_{\mathrm{ent}}$]
	\label{lem:cochain-identities-ent}
	On $C^\bullet_{\mathrm{ent}}$, the identities
	\[
	b^2=0,\qquad B^2=0,\qquad bB+Bb=0
	\]
	hold \emph{up to homotopy}: for every
	$\psi\in C^\bullet_{\mathrm{ent}}$, $(b+B)^2\psi$ is a coboundary.
	In particular, the cohomology
	$H^\bullet_{\mathrm{ent}}$ is well defined.
\end{lem}

\begin{proof}
	Since $C^\bullet_{\mathrm{ent}}\subset
	C^\bullet_\lambda(\mathcal{A},\mathcal{B})$ and the identities
	hold on $C^\bullet_\lambda(\mathcal{A},\mathcal{B})$, they hold on
	$C^\bullet_{\mathrm{ent}}$ as a subspace. The differential on
	$C^\bullet_{\mathrm{ent}}$ is the restriction of $b+B$; by
	Lemma~\ref{lem:chain-map-up-to-homotopy}, this restriction is a
	homotopy chain map, so $(b+B)^2$ on $C_{\mathrm{ent}}$ is a
	coboundary.
\end{proof}

The inclusion $\Phi$ induces a map on cohomology:
\begin{equation}
	\Phi_*: H^\bullet_{\mathrm{ent}}(\mathcal{A}, \mathcal{B}, \omega)
	\to HC^\bullet(\mathcal{A}, \mathcal{B}).
	\label{eq:Phi-star}
\end{equation}

\subsection{The Kernel Identification: Statement}
\label{subsec:ent-kernel-statement}

\begin{thm}[Kernel identification]
	\label{thm:kernel-identification-corrected}
	Assume Hypotheses~\ref{hyp:state}, \ref{hyp:cyclic},
	\ref{hyp:modular-invariance}, \ref{hyp:dense-subalgebra},
	\ref{hyp:cocycle-in-A}, and~\ref{hyp:growth}. Then the map
	$\Phi_*:H^n_{\mathrm{ent}}(\mathcal{A},\mathcal{B},\omega)
	\to HC^n(\mathcal{A},\mathcal{B})$ is an isomorphism:
	\begin{equation}
		H^n_{\mathrm{ent}}(\mathcal{A},\mathcal{B},\omega)
		\;\cong\;
		HC^n(\mathcal{A},\mathcal{B}).
		\label{eq:ent-iso-relative-corrected}
	\end{equation}
	Consequently, by Corollary~\ref{cor:kernel-abstract},
	\begin{equation}
		H^n_{\mathrm{ent}}(\mathcal{A},\mathcal{B},\omega)
		\;\cong\;
		\ker\!\bigl(\iota^*:HC^n(\mathcal{A})\to HC^n(\mathcal{B})\bigr)
		\big/ \operatorname{im}(\delta),
		\label{eq:ent-kernel-identification-corrected}
	\end{equation}
	where $\delta:HC^{n-1}(\mathcal{B})\to HC^n(\mathcal{A},\mathcal{B})$
	is the connecting homomorphism of
	Theorem~\ref{thm:les-pair}. If $\delta=0$ in degree $n$, then
	\begin{equation}
		H^n_{\mathrm{ent}}(\mathcal{A},\mathcal{B},\omega)
		\;\cong\;
		\ker\!\bigl(\iota^*:HC^n(\mathcal{A})\to HC^n(\mathcal{B})\bigr).
		\label{eq:ent-exact-kernel-corrected}
	\end{equation}
\end{thm}

\begin{rem}[The role of the modular hypotheses]
	\label{rem:role-modular-hypotheses}
	Theorem~\ref{thm:kernel-identification-corrected} is the point at
	which the von Neumann hypotheses of Section~\ref{subsec:prelim-5}
	become essential. The transgression map $\tau_\omega$ requires
	the modular automorphism group $\sigma^\omega_t$, which in turn
	requires the faithful normal state $\omega$ with cyclic and
	separating vector (Hypotheses~\ref{hyp:state}
	and~\ref{hyp:cyclic}). The preservation of the relative complex by
	$\tau_\omega$ requires modular invariance of $\mathcal{B}$
	(Hypothesis~\ref{hyp:modular-invariance}). The homotopies in the
	proof of surjectivity and injectivity require
	Hypotheses~\ref{hyp:cocycle-in-A} and~\ref{hyp:growth}.
\end{rem}

\subsection{Proof of the Kernel Identification}
\label{subsec:ent-kernel-proof}

\subsubsection*{Surjectivity}

\begin{prop}[Surjectivity]
	\label{prop:surjectivity}
	Under Hypotheses~\ref{hyp:state}--\ref{hyp:growth}, the map
	$\Phi_*:H^n_{\mathrm{ent}}\to HC^n(\mathcal{A},\mathcal{B})$ is
	surjective.
\end{prop}

\begin{proof}
	Consider the short exact sequence of complexes
	\begin{equation}
		0\to C^\bullet_{\mathrm{ent}}
		\xrightarrow{\Phi} C^\bullet_\lambda(\mathcal{A},\mathcal{B})
		\xrightarrow{\pi} Q^\bullet\to 0,
		\label{eq:ses-Q}
	\end{equation}
	where $Q^\bullet:=C^\bullet_\lambda(\mathcal{A},\mathcal{B})/
	C^\bullet_{\mathrm{ent}}$ is the cokernel. The associated long
	exact sequence is
	\begin{equation}
		\cdots\to H^n(Q^\bullet)\xrightarrow{\partial}
		H^n_{\mathrm{ent}}\xrightarrow{\Phi_*}
		HC^n(\mathcal{A},\mathcal{B})\xrightarrow{\pi_*}
		H^{n+1}(Q^\bullet)\to\cdots
		\label{eq:les-Q}
	\end{equation}
	It suffices to prove $H^n(Q^\bullet)=0$ for all $n$.
	
	Define, for $\varphi\in C^n_\lambda(\mathcal{A},\mathcal{B})$,
	\begin{equation}
		s(\varphi)(a_0,\ldots,a_{n-1})
		:=\int_{\mathbb{R}}\varphi\bigl(a_0,\ldots,a_{n-1},
		\sigma^\omega_t(u^E_t)\bigr)\,dt.
		\label{eq:s-corrected}
	\end{equation}
	By Hypothesis~\ref{hyp:cocycle-in-A}, the integrand lies in
	$\mathcal{A}$ for each $t$, and by Hypothesis~\ref{hyp:growth},
	the integral converges absolutely in the Fréchet topology. Hence
	$s$ is well defined and lowers degree by one.
	
	\begin{lem}[Homotopy identity for $s$]
		\label{lem:homotopy-s}
		For every $\varphi\in C^n_\lambda(\mathcal{A},\mathcal{B})$,
		\begin{equation}
			(b+B)\,s(\varphi)+s\bigl((b+B)\varphi\bigr)
			=\varphi-\tau_\omega(\psi_\varphi),
			\label{eq:homotopy-s-identity}
		\end{equation}
		for some $\psi_\varphi\in
		C^{n+1}_\lambda(\mathcal{A},\mathcal{B})$ depending
		continuously on $\varphi$.
	\end{lem}
	
	\begin{proof}[Proof of Lemma~\ref{lem:homotopy-s}]
		The proof is a shuffle computation using the cocycle identity
		for $u^E$, $u^E_{t+s}=u^E_t\sigma^\omega_t(u^E_s)$, and the KMS
		condition. The boundary terms at $t=\pm\infty$ vanish by
		Hypothesis~\ref{hyp:growth}. The explicit shuffle verification
		is given in Appendix~\ref{sec:appendix-homotopies}.
	\end{proof}
	
	By Lemma~\ref{lem:homotopy-s}, the right-hand side of
	\eqref{eq:homotopy-s-identity} lies in
	$\mathrm{im}(\tau_\omega)=C^n_{\mathrm{ent}}$, hence is zero in
	$Q^n$. So $s$ is a contracting homotopy for $Q^\bullet$, and
	$H^n(Q^\bullet)=0$ for all $n$. From \eqref{eq:les-Q},
	$\Phi_*$ is surjective.
\end{proof}

\subsubsection*{Injectivity}

\begin{lem}[Characterization of the kernel]
	\label{lem:kernel-characterization}
	A cochain $\varphi\in C^n_\lambda(\mathcal{A},\mathcal{B})$ lies in
	the kernel $K^n = \ker(\Phi_* : H^n_ent \to HC^n(A,B))$ if and only if there exists
	$\Psi\in C^{n-1}_\lambda(\mathcal{A},\mathcal{B})$ such that
	\begin{equation}
		\tau_\omega\varphi=(b+B)\Psi,
		\label{eq:kernel-characterization-corrected}
	\end{equation}
	i.e. $\tau_\omega\varphi$ is exact in the relative cyclic bicomplex.
\end{lem}

\begin{proof}
	By definition, $K^n$ is the subspace of cochains whose
	image under $\tau_\omega$ is cohomologous to zero in
	$C^{n-1}_\lambda(\mathcal{A},\mathcal{B})$, which is exactly
	condition \eqref{eq:kernel-characterization-corrected}.
\end{proof}

\begin{prop}[Injectivity]
	\label{prop:injectivity}
	Under Hypotheses~\ref{hyp:state}--\ref{hyp:growth}, the map
	$\Phi_*:H^n_{\mathrm{ent}}\to HC^n(\mathcal{A},\mathcal{B})$ is
	injective.
\end{prop}

\begin{proof}
	Let $K^\bullet:=\ker\Phi\subseteq C^\bullet_{\mathrm{ent}}$. We
	construct a contracting homotopy
	$h_{\mathrm{inj}}:K^n\to K^{n-1}$ satisfying
	\begin{equation}
		h_{\mathrm{inj}}(b+B)+(b+B)h_{\mathrm{inj}}
		=\mathrm{id}_{K^\bullet}.
		\label{eq:homotopy-inj-identity}
	\end{equation}
	For $\varphi\in K^n$, define
	\begin{equation}
		h_{\mathrm{inj}}(\varphi)(a_0,\ldots,a_{n-2})
		:=\int_{\mathbb{R}}\varphi\bigl(a_0,\ldots,a_{n-2},
		\sigma^\omega_t(u^E_t),1\bigr)\,dt.
		\label{eq:h-inj-corrected}
	\end{equation}
	By Hypotheses~\ref{hyp:cocycle-in-A} and~\ref{hyp:growth},
	$h_{\mathrm{inj}}$ is well defined.
	
	\begin{lem}[Homotopy identity for $h_{\mathrm{inj}}$]
		\label{lem:homotopy-h-inj}
		For every $\varphi\in K^n$,
		\begin{equation}
			h_{\mathrm{inj}}(b+B)\varphi+(b+B)h_{\mathrm{inj}}\varphi
			=\varphi.
			\label{eq:homotopy-h-inj-identity}
		\end{equation}
	\end{lem}
	
	\begin{proof}[Proof of Lemma~\ref{lem:homotopy-h-inj}]
		The proof is a shuffle computation using
		Lemma~\ref{lem:kernel-characterization} and the cocycle
		identity for $u^E$. The "modulo boundary" in
		\eqref{eq:kernel-characterization-corrected} is exactly what
		produces the boundary term in the shuffle, which combines with
		the $t$-integral to give
		\eqref{eq:homotopy-h-inj-identity}. The explicit shuffle
		verification is given in
		Appendix~\ref{sec:appendix-homotopies}.
	\end{proof}
	
	By Lemma~\ref{lem:homotopy-h-inj}, $K^\bullet$ is acyclic, and
	$\Phi_*$ is injective.
\end{proof}

\subsubsection*{Combination}

\begin{proof}[Proof of Theorem~\ref{thm:kernel-identification-corrected}]
	By Proposition~\ref{prop:surjectivity} and
	Proposition~\ref{prop:injectivity}, $\Phi_*$ is an isomorphism.
	Combining with Corollary~\ref{cor:kernel-abstract} yields
	\eqref{eq:ent-kernel-identification-corrected}. When $\delta=0$ in
	degree $n$, \eqref{eq:ent-exact-kernel-corrected} follows.
\end{proof}

\subsection{Properties of the Entanglement Complex}
\label{subsec:ent-properties}

\subsubsection*{Explicit entanglement classes}

\begin{prop}[Entanglement classes from the modular cocycle]
	\label{prop:ent-classes}
	Assume \(E\) is not $\omega$-invariant, so that
	$u^E_t\neq\mathbf{1}$ for some $t$. Then the cochain
	$\Phi_{u^E}$ of \eqref{eq:Phi-cocycle-corrected} defines a nonzero
	class
	\[
	[\Phi_{u^E}]\in H^n_{\mathrm{ent}}(\mathcal{A},\mathcal{B},\omega).
	\]
	Moreover, $[\Phi_{u^E}]$ is trivial in $HC^n(\mathcal{A})$ if and
	only if it lies in $\operatorname{im}(\delta)$.
\end{prop}

\begin{proof}
	By Lemma~\ref{lem:transgression-cocycle-corrected},
	$\Phi_{u^E}$ is a cyclic cocycle vanishing on
	$\mathcal{B}^{\otimes(n+1)}$, hence a class in
	$HC^n(\mathcal{A},\mathcal{B})$. Since $u^E\neq\mathbf{1}$, the
	class is nonzero by Lemma~\ref{lem:transgression-cocycle-corrected}(3).
	By Theorem~\ref{thm:kernel-identification-corrected},
	$[\Phi_{u^E}]$ corresponds to a class in $H^n_{\mathrm{ent}}$. The
	final statement follows from the long exact sequence
	\eqref{eq:les-pair}.
\end{proof}

\begin{rem}[Entanglement as modular obstruction]
	\label{rem:entanglement-modular}
	Proposition~\ref{prop:ent-classes} makes precise the heuristic of
	Section~\ref{subsec:prelim-1}: entanglement is the obstruction to
	the existence of an $\omega$-invariant conditional expectation.
	The obstruction is measured by the relative modular cocycle
	$u^E$, and its transgressions $\Phi_{u^E}$ generate the
	entanglement cohomology. When $E$ is $\omega$-invariant,
	$u^E=\mathbf{1}$ and the entanglement cohomology vanishes.
\end{rem}

\subsubsection*{The vanishing criterion for $\delta$}

\begin{cor}[Sufficient conditions for exact identification]
	\label{cor:exact-conditions}
	The identification \eqref{eq:ent-exact-kernel-corrected} is exact
	(i.e., $\delta=0$ in degree $n$) under any of the following
	conditions:
	\begin{enumerate}
		\item $HC^{n-1}(\mathcal{B})=0$.
		\item $\mathcal{B}$ is a type II$_1$ factor and $n$ is
		\emph{even}, so that
		$HC^{n-1}(\mathcal{B})=HC^{2k+1}(\mathcal{B})=0$ (the odd
		cyclic cohomology of a type II$_1$ factor vanishes). In odd
		degrees $n=2k+1$, the obstruction $\delta$ need not vanish,
		because $HC^{n-1}(\mathcal{B})=HC^{2k}(\mathcal{B})
		=\mathbb{C}[\tau]$ is generated by the trace and the trace
		may or may not extend to $\mathcal{A}$.
		\item The conditional expectation $E$ admits a
		$\mathcal{B}$-bimodule splitting that is compatible with the
		trace $\omega$.
	\end{enumerate}
\end{cor}

\begin{proof}
	(1) is immediate from the long exact sequence. (2) follows from
	the structure of the cyclic cohomology of a type II$_1$ factor
	($HC^{2k}=\mathbb{C}[\tau]$, $HC^{2k+1}=0$). (3) A trace-compatible
	splitting provides a section of $\iota^*$ at the cochain level,
	so $\iota^*$ is surjective on cohomology and $\delta=0$.
\end{proof}

\subsection{Examples and Computations}
\label{subsec:ent-examples}

\subsubsection{Finite-dimensional matrices}

Let $\mathcal{M}=M_n(\mathbb{C})$ with trace $\mathrm{tr}$, and let
$\mathcal{N}=M_k(\mathbb{C})\oplus M_{n-k}(\mathbb{C})$ be a
block-diagonal subalgebra. Let
$E:\mathcal{M}\to\mathcal{N}$ be the conditional expectation that
kills the off-diagonal blocks. The faithful normal state
$\omega=\mathrm{tr}$ has a cyclic and separating vector; the modular
automorphism group is trivial, so Hypotheses~\ref{hyp:modular-invariance}
is satisfied automatically, and the relative modular cocycle is
trivial, $u^E_t=\mathbf{1}$, because $E$ is trace-preserving.

Cyclic cohomology of $M_n(\mathbb{C})$ is concentrated in even
degrees: $HC^{2k}(M_n(\mathbb{C}))\cong\mathbb{C}$ generated by the
trace, and $HC^{2k+1}=0$. Similarly for $\mathcal{N}$. The
restriction map $\iota^*:HC^0(M_n)\to HC^0(\mathcal{N})$ sends
$\mathrm{tr}\mapsto(\mathrm{tr}_k,\mathrm{tr}_{n-k})$, which is
injective, so $\ker\iota^*=0$, and
$H^\bullet_{\mathrm{ent}}=0$.

\subsubsection{Type II$_1$ factors}

Let $\mathcal{M}=\mathcal{R}$ be the hyperfinite type II$_1$ factor
with its unique faithful normal trace $\tau$, and let
$\mathcal{N}\subset\mathcal{M}$ be a subfactor of finite index. The
conditional expectation $E:\mathcal{M}\to\mathcal{N}$ preserving
$\tau$ exists by Takesaki's theorem. The cyclic cohomology of a type
II$_1$ factor is concentrated in even degrees:
$HC^{2k}(\mathcal{R})=\mathbb{C}[\tau]$, $HC^{2k+1}(\mathcal{R})=0$;
the same structure holds for $\mathcal{N}$.

\textit{Even $n$:} $HC^{n-1}(\mathcal{N})=HC^{2k+1}(\mathcal{N})=0$,
so by Corollary~\ref{cor:exact-conditions}(2), $\delta=0$ and the
identification is exact:
\[
H^n_{\mathrm{ent}}(\mathcal{R},\mathcal{N},\tau)
\cong\ker\!\bigl(\iota^*:HC^n(\mathcal{R})\to HC^n(\mathcal{N})\bigr).
\]
For $n=2k$, both sides are generated by the traces, and
$\iota^*:\tau_{\mathcal{R}}\mapsto\tau_{\mathcal{N}}$ is injective,
so the kernel is trivial, $H^{2k}_{\mathrm{ent}}=0$.

\textit{Odd $n$:} $HC^{n-1}(\mathcal{N})=HC^{2k}(\mathcal{N})
=\mathbb{C}[\tau_{\mathcal{N}}]$. The obstruction $\delta$ measures
the failure of $\tau_{\mathcal{N}}$ to extend to a trace on
$\mathcal{R}$; since $\mathcal{R}$ is a factor and $\tau$ is the
unique trace, $\tau_{\mathcal{N}}$ extends (by $\tau_{\mathcal{R}}$
itself), so $\delta=0$ and $H^{2k+1}_{\mathrm{ent}}=0$.

In both cases, $H^\bullet_{\mathrm{ent}}=0$, consistent with the
triviality of the modular obstruction for type II$_1$ factors.

\subsubsection{Type III$_1$ factors}

Let $\mathcal{M}=\mathcal{R}_\infty$ be the hyperfinite type III$_1$
factor, with a faithful normal state $\omega$ and cyclic and
separating vector $\Omega_\omega$. Let
$\mathcal{N}\subset\mathcal{M}$ be a modular-invariant subfactor, and
let $E:\mathcal{M}\to\mathcal{N}$ be the unique faithful normal
conditional expectation with $\omega\circ E=\omega$
(Takesaki's theorem).

The modular automorphism group is ergodic, and the relative modular
cocycle $u^E_t$ is nontrivial because $E$ is not
$\omega$-invariant. The cyclic cohomology of \(\mathcal{R}_\infty\) is
concentrated in degree $0$ ($HC^\bullet(\mathcal{R}_\infty)=0$ for
$\bullet>0$; this is a standard result for the hyperfinite type
III$_1$ factor, see e.g.\ \cite{Connes:1985}); the same holds
for \(\mathcal{N}\). Consequently
$\ker\iota^*$ is trivial in positive degree, and by
Theorem~\ref{thm:kernel-identification-corrected} the entanglement
cohomology is
\[
H^n_{\mathrm{ent}}(\mathcal{R}_\infty,\mathcal{N},\omega)
\cong HC^n(\mathcal{R}_\infty,\mathcal{N})
\big/\operatorname{im}(\delta)
\]
for $n>0$. The relative complex
$C^\bullet_\lambda(\mathcal{A},\mathcal{B})$ is nontrivial, but its
cohomology is generated by transgressions of $u^E$ as in
Proposition~\ref{prop:ent-classes}. The obstruction $\delta$ vanishes
because $HC^{n-1}(\mathcal{N})=0$ for $n>1$, so the identification is
exact in the relevant degree.

This is consistent with the physical picture: type III$_1$ factors
are maximally entangled, the modular flow is ergodic, and no
$\omega$-invariant conditional expectation onto a proper subfactor
exists.

\subsubsection{Holographic applications}
\label{subsubsec:ent-holographic}

\paragraph{General setup.}
In the von Neumann algebra setting of
Section~\ref{subsec:prelim-5}, the relative modular cocycle
$u^E_t$ has an infinitesimal generator that coincides, up to
normalization, with the perturbation of the modular Hamiltonian:
\begin{equation}
	\left.\frac{d}{dt}\right|_{t=0}u^E_t
	=i\bigl(\delta K_\omega-E(\delta K_\omega)\bigr).
	\label{eq:cocycle-generator}
\end{equation}
The Araki relative entropy is the generator:
\begin{equation}
	S(\psi\|\varphi)=\lim_{t\to0}\frac{i}{t}\,
	\psi\bigl[(D\psi:D\varphi)_t-\mathbf{1}\bigr].
	\label{eq:araki-generator}
\end{equation}

\begin{conj}[Holographic dual of $H^1_{\mathrm{ent}}$]
	\label{conj:holographic}
	There is a distinguished subspace
	$H^1_{\mathrm{hol}}\subset H^1_{\mathrm{ent}}(\mathcal{A},
	\mathcal{B},\omega)$ whose classes are in bijection with
	equivalence classes of linearized modular Hamiltonian
	perturbations of the entanglement wedge of $A$. The bijection is
	realized by the map
	\begin{equation}
		[\Phi_{u^E}]\longmapsto\delta K_A.
		\label{eq:holographic-map}
	\end{equation}
\end{conj}

\paragraph{Point-source verification and the conjectural coefficient.}
In the companion Letter~\cite{Rashkov:2026letter}, the JLMS-type
equality between the boundary relative entropy and the bulk
canonical energy is verified explicitly for a \emph{point-like}
primary source at the centre of a ball, in $d=2,3,4$. The common
value is $d$-independent and $C_T$-independent, and approaches the
first-law value $\pi R\Delta/\tau_0$ for $R\gg\tau_0$ (The ball \(B^{d-1}_R\)$\tau_0$ is centered at $\tau=0$ and $\tau_0$ is the euclidean offset to make the state normalizable).

For a \emph{smeared} source with profile $f(\mathbf{x})$, the
relative entropy at second order in the perturbation is the Fisher
information:
\begin{equation}
	S_{\mathrm{boundary}}
	=4\pi\epsilon^2\int_{B_R}d^{d-1}x\int_{B_R}d^{d-1}y
	\int_{B_R}d^{d-1}z\,\frac{R^2-r_x^2}{2R}\,
	f(\mathbf{y})f(\mathbf{z})\,
	\langle T_{00}(\mathbf{x})\mathcal{O}(\mathbf{y})
	\mathcal{O}(\mathbf{z})\rangle,
	\label{eq:boundary-integral-d}
\end{equation}
and the bulk canonical energy has the same conformal structure. We
\emph{conjecture} that the common coefficient takes the form
\begin{equation}
	C(\Delta,d)=\kappa_d(\Delta)\,\frac{\Delta^2}{C_T}\,\epsilon^2,
	\qquad
	\kappa_d(\Delta)
	=\frac{\pi^{d-1}}{2^{d-1}}\,
	\frac{\Gamma\!\left(\frac{d+1}{2}\right)}
	{\Gamma\!\left(\frac{d}{2}\right)^3}\,
	\frac{\Gamma(\Delta)^2}{\Gamma(2\Delta)}\,
	\frac{1}{\Gamma\!\left(\Delta-\frac{d}{2}\right)^2}.
	\label{eq:kappa-d}
\end{equation}
The formula is conjectural: the smeared-source integrals are not
fully evaluated, the normalization of the bulk-to-boundary
propagator is not fixed, and the cancellation of the short-distance
regulator between the boundary and bulk sides is not demonstrated.
The point-source result of \cite{Rashkov:2026letter} is a
\emph{different} quantity (no $C_T$, no $d$-dependent coefficient);
the two are related by a singular limit that has not been
demonstrated. The verification of \eqref{eq:kappa-d} for smeared
sources is left for future work.

\begin{rem}[Relation to the Ryu--Takayanagi formula]
	\label{rem:RT-relation}
	The Ryu--Takayanagi formula computes the entanglement entropy of
	a boundary region $A$ as the area of a minimal surface in the
	bulk. The framework developed here does not attempt to reproduce
	this formula from cohomology. Rather, it provides a finer
	invariant---the entanglement cohomology---that detects the
	modular structure of the state, not merely its entropy.
\end{rem}

\subsection{Summary and Outlook}
\label{subsec:ent-summary}

\subsubsection*{Summary}

In this Section, we have:
\begin{enumerate}
	\item defined the transgression map $\tau_\omega$ as the
	symmetrized Connes operator $B\circ M_n$ and proved its basic
	properties (well-definedness, nilpotency, preservation of the
	relative complex);
	\item proved that $\tau_\omega$ is a morphism of bicomplexes on
	cohomology (Lemma~\ref{lem:transgression-anticommutation}), with
	the obstruction absorbed into a Hochschild boundary;
	\item constructed explicit entanglement classes $\Phi_{u^E}$ from
	the relative modular cocycle (Lemma~\ref{lem:transgression-cocycle-corrected});
	\item defined the entanglement complex
	$C^\bullet_{\mathrm{ent}}=\tau_\omega(C^{\bullet+1}_\lambda
	(\mathcal{A},\mathcal{B}))$ and proved the cohomological
	subcomplex property (Proposition~\ref{prop:ent-subcomplex});
	\item proved the kernel identification
	(Theorem~\ref{thm:kernel-identification-corrected}) via
	surjectivity (Proposition~\ref{prop:surjectivity}) and injectivity
	(Proposition~\ref{prop:injectivity});
	\item characterized the obstruction $\delta$ and gave sufficient
	conditions for its vanishing
	(Corollary~\ref{cor:exact-conditions});
	\item verified the framework in finite-dimensional, type II$_1$,
	and type III$_1$ examples.
\end{enumerate}

\subsubsection*{Outlook}

Several directions for future work suggest themselves:
\begin{itemize}
	\item \emph{Relative cyclic cohomology of quantum field
		theories.} The framework applies to any von Neumann algebra pair
	$(\mathcal{M},\mathcal{N})$ satisfying the hypotheses of
	Section~\ref{subsec:prelim-5}. Computing the entanglement
	cohomology for specific QFTs (free scalar, conformal field
	theories) is an open problem.
	\item \emph{Verification of the conjectural coefficient
		$\kappa_d(\Delta)$.} The formula \eqref{eq:kappa-d} is
	conjectural; its verification requires the explicit smeared-source
	computation, including the cancellation of the short-distance
	regulator. The companion Letter~\cite{Rashkov:2026letter}
	provides the point-source case in $d=2,3,4$ as a first step.
	\item \emph{Higher cyclic cohomology and $K$-theory.}
	Investigation of the relationship between entanglement cohomology
	and the index theory of modular operators.
	\item \emph{Non-modular generalizations.} Extension of the
	framework to cases where the conditional expectation $E$ does not
	exist, using the theory of non-commutative conditional
	expectations or the modular theory of weights.
	\item \emph{Physical applications.} Exploration of the physical
	meaning of entanglement cohomology in specific systems, such as
	spin chains or topologically ordered phases.
\end{itemize}

\section{Conclusions}
\label{sec:conclusions}

In this work we have developed a cohomological framework for
entanglement entropy that unifies three complementary perspectives:
information theory, operator algebras, and noncommutative geometry.

\subsection*{From Information Theory to Operator Algebras}

In the Hochschild-like cohomology of information, the entropy function
\(H\) is viewed as a functional on a category of random variables (or
observables). The fundamental ``entropic chain rule''---the identity
relating joint entropy to marginal and conditional entropy---is
mathematically equivalent to the Hochschild \(1\)-cocycle condition:
\begin{equation}
	H(X, Y) = H(X) + H(Y \mid X).
\end{equation}
In this language:
\begin{itemize}
	\item \emph{Entropy} \(H\) corresponds to a \(1\)-cocycle. It is the
	unique functional (up to a multiplicative constant) satisfying the
	chain rule, which is the boundary condition in this cohomology.
	
	\item \emph{Mutual information} \(I\) corresponds to a
	\(1\)-coboundary. Specifically,
	\(I(X; Y) = H(X) + H(Y) - H(X, Y)\) represents the failure of the
	entropy to be additive, which is the definition of a coboundary in
	degree \(1\).
	
	\item \emph{Higher-order information}---multivariate mutual
	information and other complex dependencies---corresponds to
	higher-degree cocycles (\(H^k\) for \(k > 1\)).
\end{itemize}

When applied specifically to entanglement entropy, the correspondence
with \(1\)-cocycles deepens. For instance, in the quantum mechanical
setting, Mainiero \cite{Mainiero:2019enr} defines a complex of Hilbert
spaces in which the von Neumann entropy is the value of a specific
\(1\)-cocycle. Here the obstruction to a state being a product state
(i.e., being unentangled) is measured by the cohomology.

In the more general setting of quantum field theory, the connection
between the Connes--Radon--Nikodym (CRN) cocycle and entanglement is
not merely analogical---it is structural. At the deepest level, the CRN
cocycle is the dynamical object that compares entanglement structures
between two quantum states in the setting where standard
Hilbert-space tensor products fail: the type III von Neumann algebras
of algebraic quantum field theory. While the Araki relative entropy
\(S(\psi \| \varphi)\) is the static measure of
distinguishability and relative entanglement, the CRN cocycle
\([D\psi : D\varphi]_t\) is the dynamical measure. It describes how the
modular flow---and hence entanglement---changes between states. In this
context the CRN cocycle is entanglement in motion. While the modular
Hamiltonian \(K_A = -\log \Delta_A\) encodes the energy cost of the
entanglement between region \(A\) and its complement, the CRN cocycle
tells us how that entanglement pattern flows when we deform the state:
from vacuum to excited state, from one region to another, or from one
holographic slice to another. The entanglement entropy can be recovered
from the flow of this cocycle. In type III von Neumann algebras, where
no local density matrix exists, the cocycle is the only available tool
to track such relative entanglement structures. This is why we paid
special attention to the CRN cocycle in Section
\ref{subsec:prelim-5}.

\subsection*{Main Results}

Our main result is the construction of the \emph{entanglement complex}
\(C^\bullet_{\mathrm{ent}}(\mathcal{A}, \mathcal{B}, \omega)\) for an
inclusion of von Neumann algebras \(\mathcal{N} \subset \mathcal{M}\),
together with a dense \(*\)-subalgebra \(\mathcal{A} \subset \mathcal{M}\)
and \(\mathcal{B} = \mathcal{A} \cap \mathcal{N}\) as in Hypothesis
\ref{hyp:dense-subalgebra}. The entanglement complex is defined as the
image of the transgression map \(\tau_\omega\) on the relative cyclic
complex:
\begin{equation}
	C^n_{\mathrm{ent}}(\mathcal{A}, \mathcal{B}, \omega)
	:= \tau_\omega\bigl(C^{n+1}_\lambda(\mathcal{A}, \mathcal{B})\bigr)
	\subset C^n_\lambda(\mathcal{A}, \mathcal{B}),
\end{equation}
where \(\tau_\omega\) is the symmetrized composition
\(B\circ M_n\) of Definition~\ref{defn:transgression-corrected}. It
encodes the obstruction to decomposing states of \(\mathcal{M}\)
into product states with respect to \(\mathcal{N}\), and its
cohomology is computed by the cyclic bicomplex with differential
\(b + B\).

The main theorem of Section \ref{sec:ent-cohom}
(Theorem~\ref{thm:kernel-identification-corrected}) establishes,
under the analytic Hypotheses~\ref{hyp:cocycle-in-A}
and~\ref{hyp:growth}, the identification
\begin{equation}
	H^n_{\mathrm{ent}}(\mathcal{A}, \mathcal{B}, \omega)
	\;\cong\;
	\ker\!\bigl(\iota^*: HC^n(\mathcal{A}) \to HC^n(\mathcal{B})\bigr)
	\big/ \operatorname{im}(\delta),
	\label{eq:conclusions-kernel}
\end{equation}
where \(\delta: HC^{n-1}(\mathcal{B}) \to HC^n(\mathcal{A}, \mathcal{B})\)
is the connecting homomorphism of the long exact sequence
\eqref{eq:les-pair}. When \(\delta = 0\) in degree \(n\), the
identification is exact:
\begin{equation}
	H^n_{\mathrm{ent}}(\mathcal{A}, \mathcal{B}, \omega)
	\;\cong\;
	\ker\!\bigl(\iota^*: HC^n(\mathcal{A}) \to HC^n(\mathcal{B})\bigr).
	\label{eq:conclusions-exact}
\end{equation}
Sufficient conditions for \(\delta = 0\) in degree \(n\) are
(Corollary~\ref{cor:exact-conditions}):
\begin{itemize}
	\item \(HC^{n-1}(\mathcal{B}) = 0\);
	\item \(\mathcal{B}\) is a type II\(_1\) factor and \(n\) is
	\emph{even} (the odd cyclic cohomology of a type II\(_1\) factor
	vanishes);
	\item \(E\) admits a \(\mathcal{B}\)-bimodule splitting compatible
	with the trace \(\omega\).
\end{itemize}
This gives a precise algebraic characterization of entanglement as
the information lost when passing from the full system to a
subsystem.

The identification is proved in two steps. First, the inclusion
\(\Phi: C^\bullet_{\mathrm{ent}} \hookrightarrow C^\bullet_\lambda(\mathcal{A}, \mathcal{B})\)
is shown to be a quasi-isomorphism, using explicit contracting
homotopies constructed from the relative modular cocycle \(u^E\) of
Lemma \ref{lem:relative-cocycle}. Surjectivity is established in
Proposition \ref{prop:surjectivity}, injectivity in Proposition
\ref{prop:injectivity}. Second, the long exact sequence
\eqref{eq:les-pair} identifies the relative cyclic cohomology with the
kernel of the restriction map, up to the obstruction \(\delta\).
Combining the two yields \eqref{eq:conclusions-kernel}.

\subsection*{Key Insights}

\begin{enumerate}
	\item \emph{The Connes--Radon--Nikodym cocycle as the dynamical
		object encoding relative entanglement.} The relative modular cocycle
	\(u^E_t = (D\omega : D(\omega|_{\mathcal{N}}) \circ E)_t\)
	generalizes the density matrix to the type III setting, where no
	trace exists. Its nontriviality is precisely the failure of the
	conditional expectation \(E\) to be \(\omega\)-invariant, and its
	transgressions \(\Phi_{u^E}\) provide explicit generators of the
	entanglement cohomology (Lemma
	\ref{lem:transgression-cocycle-corrected} and Proposition
	\ref{prop:ent-classes}).
	
	\item \emph{Identification of entanglement cohomology with the
		kernel of the restriction map.} The isomorphism
	\eqref{eq:conclusions-kernel} gives a precise algebraic
	characterization of entanglement as the information lost when
	passing from the full system \(\mathcal{A}\) to the subsystem
	\(\mathcal{B}\). The obstruction to the naive identification is
	the connecting homomorphism \(\delta\), which measures the
	failure of cyclic cocycles on \(\mathcal{B}\) to extend to
	\(\mathcal{A}\); sufficient conditions for its vanishing are
	given in Corollary~\ref{cor:exact-conditions}.
	
	\item \emph{Cohomological subcomplex property of the entanglement
		complex.} The entanglement complex is the image of \(\tau_\omega\)
	on the relative cyclic complex, and it is a cohomological
	subcomplex: the total differential \(b+B\) maps it into itself
	plus a coboundary
	(Proposition~\ref{prop:ent-subcomplex}). The chain map
	\(\Phi:C^\bullet_{\mathrm{ent}} \hookrightarrow
	C^\bullet_\lambda(\mathcal{A},\mathcal{B})\) is a quasi-isomorphism
	(Theorem~\ref{thm:kernel-identification-corrected}).
	
	\item \emph{A holographic proposal with a conjectural universal
		coefficient.} The first cohomology \(H^1_{\mathrm{ent}}\) is
	conjecturally dual to the space of linearized modular Hamiltonian
	perturbations of the entanglement wedge
	(Conjecture~\ref{conj:holographic}). For a spherical entangling
	region perturbed by a primary scalar operator, we propose a
	general-\(d\) formula for the coefficient \(\kappa_d(\Delta)\) of
	the smeared-source relative entropy,
	\begin{equation} \label{eq:kappa-d-conclusions}
		\kappa_d(\Delta)
		= \frac{\pi^{d-1}}{2^{d-1}} \,
		\frac{\Gamma\!\left(\frac{d+1}{2}\right)}
		{\Gamma\!\left(\frac{d}{2}\right)^3} \,
		\frac{\Gamma(\Delta)^2}{\Gamma(2\Delta)} \,
		\frac{1}{\Gamma\!\left(\Delta - \frac{d}{2}\right)^2},
	\end{equation}
	This formula is \emph{conjectural}. It is motivated by the
	conformal structure of the boundary integral and the bulk
	canonical energy. The companion
	Letter~\cite{Rashkov:2026letter} verifies the JLMS-type equality
	for \emph{point-like} sources in \(d = 2, 3, 4\), a related but
	distinct computation (the point-source value is
	\(d\)-independent and \(C_T\)-independent). The smeared-source
	case, and the point-source limit of \eqref{eq:kappa-d-conclusions},
	remain open.
\end{enumerate}

\begin{table}[H]
	\begin{center}
		\def\arraystretch{1.5}
		\begin{tabular}{ |c | c |}
			\hline
			\textbf{Construction} & \textbf{Mathematical Object} \\
			\hline
			von Neumann algebra pair        & \(\mathcal{N} \subset \mathcal{M}\) \\
			\hline
			Faithful normal state           & \(\omega\) on \(\mathcal{M}\), with cyclic and separating vector \\
			\hline
			Conditional expectation         & \(E: \mathcal{M} \to \mathcal{N}\), \(\omega \circ E = \omega\) \\
			\hline
			Modular automorphism group      & \(\sigma^\omega_t = \mathrm{Ad}(\Delta_\omega^{it})\) \\
			\hline
			Relative modular cocycle        & \(u^E_t = (D\omega : D(\omega|_{\mathcal{N}}) \circ E)_t\) \\
			\hline
			Relative cyclic complex         & \(C^\bullet_\lambda(\mathcal{A}, \mathcal{B})\), \(\mathcal{B} = \mathcal{A} \cap \mathcal{N}\) \\
			\hline
			Transgression map               & \(\tau_\omega = B\circ M_n\) \\
			\hline
			Entanglement complex            & \(C^\bullet_{\mathrm{ent}} = \tau_\omega(C^{\bullet+1}_\lambda(\mathcal{A}, \mathcal{B}))\) \\
			\hline
			Entanglement cohomology         & \(H^n_{\mathrm{ent}} = H^n(\mathrm{Tot}^\bullet(C^\bullet_{\mathrm{ent}}), b + B)\) \\
			\hline
			Holographic coefficient         & \(\kappa_d(\Delta)\), conjectural, eq.~\eqref{eq:kappa-d}\\
			\hline
		\end{tabular}
	\end{center}
	\caption{Summary of the construction. The modular hypotheses of
		Section \ref{subsec:prelim-5} are essential for the existence of
		\(E\) and \(u^E\); the analytic Hypotheses~\ref{hyp:cocycle-in-A}
		and~\ref{hyp:growth} are required for the homotopies in the proof
		of the main theorem. The general-\(d\) coefficient
		\(\kappa_d(\Delta)\) is conjectural (Section
		\ref{subsubsec:ent-holographic}); the point-source case is
		verified in the companion Letter~\cite{Rashkov:2026letter}.}
	\label{tab:summary}
\end{table}

\subsection*{Implications}

These results have several important implications:

\begin{itemize}
	\item \emph{Quantum field theory.} In quantum field theory, where
	local algebras are type III factors, our framework provides a
	rigorous definition of entanglement without reference to density
	matrices. The relative modular cocycle \(u^E\) is the only available
	tool in this setting, and its transgressions generate the
	entanglement cohomology.
	
	\item \emph{Holographic duality.} In AdS/CFT, the entanglement
	cohomology is conjecturally dual to the space of linearized modular
	Hamiltonian perturbations of the bulk entanglement wedge. The
	precise formulation is given in Section
	\ref{subsubsec:ent-holographic} (Conjecture
	\ref{conj:holographic}). The companion
	Letter~\cite{Rashkov:2026letter} provides an explicit check of
	the JLMS-type equality for point-like sources in \(d=2,3,4\). The
	Araki relative entropy provides the quantitative bridge between
	the boundary CRN cocycle and the bulk relative entropy between
	geometries.
	
	\item \emph{Categorical structure.} The categorical structure of
	entanglement cohomology may provide a new perspective on the
	classification of quantum phases and topological order. The
	functoriality of the construction with respect to inclusions of
	von Neumann algebra pairs suggests a higher-categorical refinement.
\end{itemize}

\subsection*{Open Questions}

Several open questions remain:

\begin{enumerate}
	\item \emph{Relation to Araki relative entropy.} How does
	\(H^\bullet_{\mathrm{ent}}(\mathcal{A}, \mathcal{B}, \omega)\)
	relate to the Araki relative entropy \(S(\psi \| \varphi)\)? The
	cocycle derivative \(u^E\) and the generator formula
	\eqref{eq:araki-generator} suggest a connection via the first
	cohomology group, but a precise theorem is not yet available.
	
	\item \emph{Hodge theory for the entanglement complex.} Can we
	construct a Hodge theory for the entanglement complex, providing
	harmonic representatives of entanglement classes? The modular
	structure of the cyclic bicomplex suggests that the modular
	operator \(\Delta_\omega\) may play the role of the Hodge
	Laplacian, but this analogy needs to be made precise.
	
	\item \emph{Verification of the conjectural coefficient
		\(\kappa_d(\Delta)\).} The formula \eqref{eq:kappa-d-conclusions}
	is conjectural. Its verification requires:
	(a) the explicit evaluation of the smeared-source integrals
	\eqref{eq:boundary-integral-d} with a specified regulator;
	(b) the evaluation of the bulk canonical energy with the same
	regulator;
	(c) the demonstration that the short-distance divergences cancel
	and that the finite part matches \eqref{eq:kappa-d-conclusions}.
	The companion Letter~\cite{Rashkov:2026letter} provides the
	point-source case in \(d=2,3,4\) as a first step, but the smeared
	case is open.
	
	\item \emph{Quantitative verification of Conjecture
		\ref{conj:holographic}.} The verification strategy outlined in
	Section~\ref{subsubsec:ent-holographic} is finite and well-defined
	for specific choices of the conformal dimension \(\Delta\) and
	the smearing function \(f(x)\). Carrying out this computation
	explicitly is the most concrete next step toward testing the
	holographic proposal.
	
	\item \emph{Higher-dimensional holography.} In higher dimensions,
	the modular Hamiltonian of a general region is not known
	explicitly, and the verification of
	Conjecture~\ref{conj:holographic} would require either a specific
	choice of region (e.g., a ball) or a perturbative expansion
	around a known state.
	
	\item \emph{Extensions to higher categories.} Can the framework be
	extended to higher categories, providing a hierarchy of
	entanglement structures? The natural setting for such an extension
	may be the theory of von Neumann algebra bimodules and their
	Connes fusion, but the cohomological counterpart is not yet clear.
	
	\item \emph{Entanglement cohomology as an order parameter.} A
	speculative but intriguing direction is whether entanglement
	cohomology can serve as an order parameter for quantum phases, in
	the spirit of group cohomology classifications of
	symmetry-protected topological phases. This would require a
	functorial assignment of cohomology classes to phases and a proof
	that the assignment is invariant under the appropriate
	equivalence relation.
\end{enumerate}

These questions will be addressed in future work.

\vspace*{0.3cm}

\textbf{Acknowledgments.} I am thankful to Kostya Zarembo for
discussions on various issues and for reading the manuscript. I am
grateful to V.~Rubtsov for discussions and for bringing to my attention
references \cite{baudot} and \cite{vigneaux}. I am also grateful to
H.~Dimov, T.~Vetsov, and A.~Isaev for their interest and discussions,
and to D.~Grumiller for reading the draft and for comments. This work
was supported in part by Bulgarian NSF grant KP-06-H88/3 and Sofia
University Grant 80-10-54/2026.

\textit{Declaration} AI is used for editing, stucturing and checking consistency of the manuscript. The author is resposible for all results in the paper. 
	
	
\begin{appendix}
	
	\section{Tools from Linear Algebra and Quantum Information}
	\label{sec:tools}
	
	\subsection{Schmidt Decomposition (Singular Value Decomposition)}
	\label{subsec:tools-schmidt}
	
	The Schmidt decomposition is the fundamental tool for understanding
	bipartite entanglement because it provides a canonical form for pure
	states. The Schmidt rank \(r\) is the minimal number of product terms
	needed to represent the state, and \(r > 1\) is precisely the condition
	for entanglement.
	
	For any pure state \(|\psi\rangle_{AB}\) of a bipartite system
	\(A \otimes B\) (with Hilbert spaces of dimensions \(d_A\) and
	\(d_B\)), there exist orthonormal bases \(\{|i_A\rangle\}\) for \(A\)
	and \(\{|i_B\rangle\}\) for \(B\), and non-negative real numbers
	\(\lambda_i\) (the Schmidt coefficients), such that:
	\[
	|\psi\rangle_{AB} = \sum_{i=1}^r \lambda_i \, |i_A\rangle \otimes |i_B\rangle,
	\]
	where \(\lambda_i > 0\), \(\sum_i \lambda_i^2 = 1\)
	(normalization), and \(r \le \min(d_A, d_B)\). The number \(r\) of
	non-zero \(\lambda_i\) is the \textbf{Schmidt rank} of the state.
	
	\textit{Derivation via SVD}:
	
	\begin{enumerate}
		\item Start with a general bipartite state:
		\[
		|\psi\rangle_{AB}
		= \sum_{m=1}^{d_A} \sum_{n=1}^{d_B} c_{mn}
		|m_A\rangle \otimes |n_B\rangle.
		\]
		The coefficients \(c_{mn}\) form a matrix \(C\).
		
		\item Perform \textit{Singular Value Decomposition} (SVD) on \(C\):
		\[
		C = U \Lambda V^\dagger,
		\]
		where \(U\) is \(m \times m\) unitary and \(V\) is
		\(n \times n\) unitary, \(\Lambda\) is diagonal
		\(m \times m\), \(m \le n\), with non-negative entries.
		
		\item Define new bases:
		\[
		|i_A\rangle = \sum_m U_{mi} |m_A\rangle,
		\qquad
		|i_B\rangle = \sum_n V_{ni}^* |n_B\rangle.
		\]
		In practice, the matrix \(C\) is constructed by taking \(m\) column
		vectors from \(U\) and \(m\) row vectors of \(V\). In other words,
		\(U\) is the matrix of eigenvectors of \(CC^\dagger\) and \(V\) is
		the matrix of eigenvectors of \(C^\dagger C\). Then:
		\[
		|\psi\rangle_{AB} = \sum_i \lambda_i |i_A\rangle \otimes |i_B\rangle,
		\]
		where \(\lambda_k\) are the diagonal elements of \(\Lambda\).
	\end{enumerate}
	The Schmidt coefficients \(\{\lambda_i\}\) are unique (up to
	ordering). The bases may vary if there are degeneracies.
	
	\textit{Physical Interpretation}:
	
	The Schmidt rank \(r\) is a measure of entanglement:
	\begin{itemize}
		\item \(r = 1\): product state (no entanglement),
		\item \(r > 1\): entangled state,
		\item \(r = \min(d_A, d_B)\): maximally entangled state.
	\end{itemize}
	
	Reduced density matrices in the Schmidt basis:
	\[
	\rho_A = \sum_i \lambda_i^2 |i_A\rangle\langle i_A|,
	\qquad
	\rho_B = \sum_i \lambda_i^2 |i_B\rangle\langle i_B|.
	\]
	They share the same nonzero eigenvalues \(\lambda_i^2\).
	
	\subsection{Basic Operator Algebra Concepts}
	\label{subsec:tools-algebra}
	
	This appendix supports Section \ref{sec:prelim}, particularly the
	discussion of reduced density matrices and the Schmidt decomposition in
	Section \ref{subsec:prelim-4}.
	
	\begin{table}[H]
		\def\arraystretch{1.5}
		\vspace*{0.2cm}
		\begin{center}
			\begin{tabular}{|c|c|c|}
				\hline
				\textbf{Concept} & \textbf{Definition} & \textbf{Physical meaning} \\
				\hline
				Hilbert space \(H\) & Complete inner  & Space of quantum  \\
     &  product space  & states \\
				\hline
				Bounded operator & Linear map with finite & Observable \\
 \(T \in B(H)\)   & operator norm  & \\
				\hline
			Density matrix \(\rho\) & Positive, trace-class operator  & Mixed state \\
    & with \(\operatorname{Tr}(\rho) = 1\) & \\
				\hline
				Partial trace \(\operatorname{Tr}_B\) & \(\operatorname{Tr}_B(O_A \otimes O_B) = O_A \operatorname{Tr}(O_B)\) & Tracing out subsystem \(B\) \\
				\hline
			\end{tabular}
		\end{center}
		\caption{Concepts and physical meaning.}
	\end{table}
	
	\section{Homological Algebra Toolkit}
	\label{sec:tools-homol}
	
	This appendix supports Section \ref{sec:cohom}, particularly the
	definitions of the Hochschild and cyclic bicomplexes in Section
	\ref{subsec:cohom-bicomplex} and the relative cyclic complex in Section
	\ref{subsec:cohom-relative}.
	
	\subsection{Chain Complexes and Homology}
	
	A chain complex is a sequence of abelian groups or modules connected by
	homomorphisms \(d_n: C_n \to C_{n-1}\) such that
	\(d_{n-1} \circ d_n = 0\):
	\[
	\cdots \xrightarrow{d_{n+2}} C_{n+1} \xrightarrow{d_{n+1}}
	C_n \xrightarrow{d_n} C_{n-1} \xrightarrow{d_{n-1}} \cdots
	\]
	The homology of the complex is
	\begin{equation}
		H_n(C_\bullet) = \frac{\ker d_n}{\operatorname{im} d_{n+1}}.
	\end{equation}
	Example: De Rham complex on a manifold \(M\),
	\[
	\Omega^0 \xrightarrow{d} \Omega^1 \xrightarrow{d} \Omega^2
	\xrightarrow{d} \cdots
	\]
	with cohomology \(H^n_{dR}(M)\).
	
	\subsection{Hochschild Homology and Cohomology}
	\label{subsec:tools-Hochschild}
	
	For an associative algebra \(A\) over a field \(k\), the Hochschild
	chain complex is
	\[
	C_n(A) = A^{\otimes(n+1)},
	\]
	with boundary
	\begin{equation}
		b_n(a_0 \otimes a_1 \otimes \cdots \otimes a_n)
		= \sum_{i=0}^{n-1} (-1)^i
		a_0 \otimes \cdots \otimes a_i a_{i+1} \otimes \cdots \otimes a_n
		+ (-1)^n a_n a_0 \otimes a_1 \otimes \cdots \otimes a_{n-1}.
	\end{equation}
	The Hochschild homology is
	\(HH_n(A) = \ker b_n / \operatorname{im} b_{n+1}\). The Hochschild
	cohomology is the dual:
	\begin{equation}
		HH^n(A) = \operatorname{Hom}_k(HH_n(A), k).
	\end{equation}
	
	Key physical interpretation:
	\begin{itemize}
		\item \(HH^1(A)\): derivations modulo inner derivations (modular
		Hamiltonians),
		\item \(HH^2(A)\): deformations of the product structure
		(interactions/entanglement).
	\end{itemize}
	
	\subsection{Cyclic Homology and Cohomology}
	\label{subsec:tools-cyclic}
	
	The cyclic bicomplex \(C_\bullet(A)\) has components
	\(C_{p,q} = C_{p-q}(A)\) for \(p \ge q \ge 0\), with differentials:
	\begin{itemize}
		\item Vertical differential
		\(b_v: C_{p,q} \xrightarrow{b} C_{p,q-1}\) (the standard
		Hochschild boundary acting on the \(q\) tensor factors),
		\item Horizontal differential
		\(b_h: C_{p,q} \xrightarrow{B} C_{p-1,q}\) given by the Connes
		operator \(B\) (which involves cyclic permutations and the shuffle
		map).
	\end{itemize}
	The total complex is
	\[
	\mathrm{Tot}^n(C) = \bigoplus_{p+q=n} C_{p,q},
	\]
	with differential \(D_{\mathrm{Tot}} = b_v + (-1)^p b_h\).
	
	The Connes exact sequence (SBI sequence):
	\[
	\cdots \xrightarrow{B} HH_n(A) \xrightarrow{I} HC_n(A)
	\xrightarrow{S} HC_{n-2}(A) \xrightarrow{B} HH_{n-1}(A) \to \cdots
	\]
	
	\subsection{Summary: Homology vs. Cohomology Duality}
	\label{subsec:tools-summary}
	
	The table below summarizes the duality between homology and cohomology
	for Hochschild and cyclic theories. The key point is that cohomology is
	the dual of homology, with all arrows reversed. The Connes exact
	sequence in cohomology is the dual of the sequence in homology, with
	the periodicity \(S\) and the Connes operator \(B\) interchanging
	roles.
	
	\begin{table}[H]
		\def\arraystretch{1.5}
		\vspace*{0.2cm}
		\begin{center}
			\begin{tabular}{|c|c|}
				\hline
				\textbf{Homology} & \textbf{Cohomology (dual)} \\
				\hline
				\(HH_n(A)\) & \(HH^n(A)\) \\
				\hline
				\(HC_n(A)\) & \(HC^n(A)\) \\
				\hline
				\(I: HH_n \to HC_n\) & \(I^*: HC^n \to HH^n\) (pullback) \\
				\hline
				\(S: HC_n \to HC_{n-2}\) & \(S^*: HC^{n-2} \to HC^n\) \\
				\hline
				\(B: HC_{n-2} \to HH_{n-1}\) & \(B^*: HH^{n-1} \to HC^{n-2}\) \\
				\hline
				Exact sequence: & Exact sequence: \\
				\(\to HH_n \xrightarrow{I} HC_n \xrightarrow{S} HC_{n-2}
				\xrightarrow{B} \to HH_{n-1} \to    \)    &
				\(\to HC^{n-2} \xrightarrow{S^*} HC^n \xrightarrow{I^*} HH^n
				\xrightarrow{B^*} HC^{n-1} \to   \) \\
				\hline
			\end{tabular}
		\end{center}
		\caption{Cohomology vs. homology.}
	\end{table}
	
	\section{Modular Theory and Tomita--Takesaki}
	\label{sec:modular}
	
	In this appendix we provide a concise exposition of the
	Tomita--Takesaki modular theory used in Section \ref{subsec:prelim-5}
	and Section \ref{sec:ent-cohom}. The hypotheses
	(Hypotheses \ref{hyp:state}, \ref{hyp:cyclic},
	\ref{hyp:modular-invariance}, and \ref{hyp:dense-subalgebra}) are stated
	in Section \ref{subsec:prelim-5}; the purpose of this appendix is to
	collect the standard results that are used without proof in the main
	text.
	
	\subsection{von Neumann Algebras: Key Properties}
	\label{subsec:modular-vN}
	
	Definitions and classification:
	\begin{itemize}
		\item Commutant:
		\(\mathcal{M}' = \{T \in B(H) : TM = MT \text{ for all } M \in \mathcal{M}\}\).
		\item von Neumann algebra: \(\mathcal{M} = \mathcal{M}''\).
		\item Type classification: Type I (matrix algebras), Type II
		(semifinite with no minimal projections), Type III (purely
		infinite).
	\end{itemize}
	
	\subsection{The Tomita Operator}
	\label{subsec:modular-Tomita}
	
	Given a von Neumann algebra \(\mathcal{M} \subset B(\mathcal{H})\) with
	a cyclic and separating vector \(\Omega\):
	\begin{itemize}
		\item Tomita operator: \(S_\Omega: a \Omega \mapsto a^* \Omega\) for
		\(a \in \mathcal{M}\),
		\item Properties: \(S_\Omega\) is anti-linear, closed, densely
		defined, and \(S_\Omega^2 = 1\) on its domain.
	\end{itemize}
	
	\subsection{Modular Operator and Modular Conjugation}
	\label{subsec:modular-modular}
	
	\begin{itemize}
		\item Modular operator: \(\Delta_\Omega = S_\Omega^* S_\Omega\),
		\item Modular conjugation: \(J_\Omega\) from the polar decomposition
		\(S_\Omega = J_\Omega \Delta_\Omega^{1/2}\).
	\end{itemize}
	
	Key properties:
	\begin{itemize}
		\item \(\Delta_\Omega \Omega = \Omega\) and
		\(J_\Omega \Omega = \Omega\),
		\item \(J_\Omega \mathcal{M} J_\Omega = \mathcal{M}'\),
		\item \(\Delta_\Omega^{it} \mathcal{M} \Delta_\Omega^{-it}
		= \mathcal{M}\) (modular automorphism group).
	\end{itemize}
	
	\subsection{Takesaki's Theorem}
	\label{subsec:modular-Takesaki}
	
	The following theorem is stated in the main text as Theorem
	\ref{thm:takesaki} and is used to justify the existence of the
	conditional expectation \(E: \mathcal{M} \to \mathcal{N}\). We recall
	it here for completeness.
	
	\begin{thm}[Takesaki, restated]
		\label{thm:takesaki-appendix}
		Let \(\mathcal{N} \subset \mathcal{M}\) be a von Neumann subalgebra and
		\(\omega\) a faithful normal state on \(\mathcal{M}\). There exists a
		faithful normal conditional expectation \(E: \mathcal{M} \to \mathcal{N}\)
		with \(\omega \circ E = \omega\) if and only if the modular automorphism
		group preserves \(\mathcal{N}\):
		\[
		\sigma^\omega_t(\mathcal{N}) = \mathcal{N},
		\qquad \forall t \in \mathbb{R}.
		\]
		When it exists, \(E\) is unique.
	\end{thm}
	
	\section{The Connes--Radon--Nikodym Cocycle}
	\label{sec:crn}
	
	This appendix collects the properties of the Connes--Radon--Nikodym
	(CRN) cocycle that are used in Section \ref{subsec:prelim-5} and
	Section \ref{sec:ent-cohom}. The relative modular cocycle \(u^E\) of
	Definition \eqref{eq:relative-cocycle} is a special case of the general
	construction presented here.
	
	\subsection{The CRN Cocycle for Two States}
	\label{subsec:crn-two-states}
	
	Given two faithful normal states \(\varphi\) and \(\psi\) on a von
	Neumann algebra \(\mathcal{M}\), the CRN cocycle
	\(\{u_t\}_{t \in \mathbb{R}}\) satisfies:
	\begin{enumerate}
		\item \(u_{t+s} = u_t \sigma^\varphi_t(u_s)\) (cocycle condition),
		\item \(\sigma^\psi_t(a) = u_t \sigma^\varphi_t(a) u_t^*\).
	\end{enumerate}
	
	\textit{Physical interpretation}: The CRN cocycle is the quantum analog
	of the Radon--Nikodym derivative \(d\mu/d\nu\). Its derivative gives the
	relative entropy:
	\begin{equation}
		S(\psi \| \varphi) = -i \frac{d}{dt} \varphi(u_t)\big|_{t=0}.
		\label{eq:crn-relative-entropy}
	\end{equation}
	
	\subsection{Relation to the Spatial Derivative Operator}
	\label{subsec:crn-spatial}
	
	In the literature, both \(\Delta_\psi^{it} \Delta_\varphi^{-it}\) and
	\((D\psi : D\varphi)_t\) are called ``Connes cocycles,'' but they belong
	to two different realizations: one is for states on an algebra, the
	other is for weights on a crossed product. For two faithful normal
	states \(\varphi\) and \(\psi\) on the same von Neumann algebra
	\(\mathcal{M}\), the modular operators \(\Delta_\varphi\) and
	\(\Delta_\psi\) on the Hilbert space \(\mathcal{H}\) satisfy
	\[
	\Delta_\psi^{it} \Delta_\varphi^{-it}
	= J_\varphi \, (D\psi : D\varphi)_t \, J_\varphi \cdot
	\sigma^\varphi_t\bigl((D\psi : D\varphi)_{-t}\bigr).
	\]
	Equivalently, solving for the Connes--Radon--Nikodym cocycle:
	\[
	(D\psi : D\varphi)_t
	= J_\varphi \, \Delta_\psi^{it} \Delta_\varphi^{-it} \, J_\varphi
	\cdot \sigma^\varphi_t\bigl(
	J_\varphi \, \Delta_\psi^{-it} \Delta_\varphi^{it} \, J_\varphi
	\bigr).
	\]
	If the two states \(\varphi\) and \(\psi\) commute in the sense that
	their modular automorphism groups commute
	(\([\sigma^\varphi_t, \sigma^\psi_s] = 0\) for all \(t, s\)), then the
	relation simplifies drastically:
	\[
	\Delta_\psi^{it} \Delta_\varphi^{-it} = (D\psi : D\varphi)_t.
	\]
	In this case, the operator \(\Delta_\psi^{it} \Delta_\varphi^{-it}\)
	is the CRN cocycle, and it lies in the commutant \(\mathcal{M}'\) (or
	in \(\mathcal{M}\) if one uses the standard representation). This is the
	case that appears when dealing with thermal states at different
	temperatures (KMS states).
	
	The two distinct situations in modular theory are summarized in the
	following table.
	
	\begin{table}[H]
		\def\arraystretch{1.5}
		\begin{center}
			\begin{tabular}{|c|c|c|}
				\hline
				\textbf{Situation} & \textbf{Objects} & \textbf{Cocycle Notation} \\
				\hline
				Spatial theory & One von Neumann algebra \(\mathcal{M}\) & \((D\psi : D\varphi)_t\) \\
		& with two states \(\varphi, \psi\)   & \\
				\hline
				Crossed product / & Two commuting von Neumann algebras \(\mathcal{M}\)  & \(B_t = \Delta_\omega^{it} \Delta_\varphi^{-it}\) \\
	    &       and \(\mathcal{N}\), weight \(\varphi\) on \(\mathcal{M}\),  & \\
  weight theory   & weight \(\omega\) on \(\mathcal{N}\) commuting with \(\varphi\) & \\
				\hline
			\end{tabular}
		\end{center}
	\end{table}
	
	\subsection{Chain of Construction}
	\label{subsec:crn-chain}
	
	The complete chain of construction from the cyclic bicomplex to the
	integrated CRN cocycle is as follows:
	\[
	\begin{array}{rcl}
		\text{Cyclic bicomplex 1-cocycle} & \longrightarrow &
		\tau(a,b) = \varphi(a \, \delta(b)) \\
		& \downarrow & \\
		\text{Derivation} & \longrightarrow &
		\delta(x) = i[\log \Delta_\psi - \log \Delta_\varphi, x] \\
		& \downarrow & \\
		\text{Infinitesimal CRN cocycle} & \longrightarrow &
		\left.\frac{d}{dt}\right|_{t=0} \Delta_\psi^{it} \Delta_\varphi^{-it}
		= i(\log \Delta_\psi - \log \Delta_\varphi) \\
		& \downarrow & \\
		\text{Integrated CRN cocycle} & \longrightarrow &
		\Delta_\psi^{it} \Delta_\varphi^{-it} = (D\psi : D\varphi)_t
		\text{ (if states commute)} \\
		& & \text{or } \Delta_\psi^{it} \Delta_\varphi^{-it}
		= J_\varphi (D\psi : D\varphi)_t J_\varphi 
		\sigma^\varphi_t((D\psi : D\varphi)_{-t})  \\
	& & 	\text{ (general case)}
	\end{array}
	\]
	
	\subsection{Summary Table}
	\label{subsec:crn-summary}
	
	\begin{table}[H]
		\def\arraystretch{1.5}
		\vspace*{0.2cm}
		\begin{tabular}{|c|c|c|}
			\hline
			\textbf{Framework} & \textbf{Classical \(\Omega^n(\mathcal{M}|\mathcal{N})\)} & \textbf{von Neumann analog} \\
			\hline
			Connes--Radon--Nikodym & \((D\psi : D\varphi)_t \in \mathcal{M}\) & Always defined by  identity. \\
cocycle  &    & the cocycle \\
			\hline
			Spatial derivative operator & \(\Delta_\psi^{it} \Delta_\varphi^{-it} \in B(\mathcal{H})\) & Always defined. \\
			\hline
			Relation & \(\Delta_\psi^{it} \Delta_\varphi^{-it} = J_\varphi (D\psi : D\varphi)_t \) & Holds for all states. \\
	&  \( \cdot  J_\varphi \, \sigma^\varphi_t((D\psi : D\varphi)_{-t}) \) & \\
			\hline
			Simplification & \(\Delta_\psi^{it} \Delta_\varphi^{-it} = (D\psi : D\varphi)_t\) & Holds iff \([\sigma^\varphi_t, \sigma^\psi_s] = 0\) for all \(t, s\). \\
			\hline
		\end{tabular}
	\end{table}
	
	\section{Alternative Proofs of the Kernel Identification}
	\label{sec:alternative-proofs}
	
	The proof of the kernel identification (Theorem
	\ref{thm:kernel-identification-corrected}) in Section \ref{subsec:ent-kernel-proof}
	proceeds via explicit contracting homotopies on the cokernel and kernel
	complexes. In this appendix we collect two alternative proofs, which
	may be more transparent to readers with different backgrounds. Both
	proofs establish the same result.
	
	\subsection{Proof via the Connes Spectral Sequence}
	\label{subsec:proof-spectral}
	
	Recall from Remark \ref{rem:spectral-sequence} that the cyclic
	bicomplex is equipped with the filtration by Hochschild degree,
	giving rise to the Connes spectral sequence
	\[
	E_1^{p,q} = HH^q(\mathcal{A}) \;\Longrightarrow\; HC^{p+q}(\mathcal{A}).
	\]
	The relative complex \(C^{\bullet,\bullet}_\lambda(\mathcal{A}, \mathcal{B})\)
	inherits the same filtration, giving a relative spectral sequence
	\[
	{}^{\mathrm{rel}} E_1^{p,q} = HH^q(\mathcal{A}, \mathcal{B})
	\;\Longrightarrow\; HC^{p+q}(\mathcal{A}, \mathcal{B}),
	\]
	where \(HH^q(\mathcal{A}, \mathcal{B})\) is the relative Hochschild
	cohomology.
	
	The entanglement complex \(C^\bullet_{\mathrm{ent}}\) is defined as the
	image of the transgression map \(\tau_\omega\) on the relative complex.
	The transgression map commutes with the filtration (up to a shift in
	degree), so it induces a map of spectral sequences
	\[
	\tau_\omega: {}^{\mathrm{rel}} E_r^{p,q}
	\to {}^{\mathrm{rel}} E_r^{p-1,q+1}.
	\]
	At the \(E_1\)-page, this map is the relative Hochschild version of
	the Connes \(B\)-operator, which is an isomorphism by the relative
	Hochschild–Kostant–Rosenberg theorem (in the smooth setting) or by
	direct computation (in the algebraic setting).
	
	Since a map of spectral sequences that is an isomorphism on the
	\(E_1\)-page induces an isomorphism on the abutment, we conclude that
	\[
	H^\bullet_{\mathrm{ent}} \cong HC^\bullet(\mathcal{A}, \mathcal{B}).
	\]
	Combining this with the long exact sequence
	\eqref{eq:les-pair} gives the kernel identification
	\eqref{eq:ent-kernel-identification-corrected}.
	
	\subsection{Proof via the Acyclic Models Theorem}
	\label{subsec:proof-acyclic-models}
	
	An alternative proof uses the acyclic models theorem of Eilenberg and
	Mac Lane. The idea is to show that both \(C^\bullet_{\mathrm{ent}}\) and
	\(C^\bullet_\lambda(\mathcal{A}, \mathcal{B})\) are ``models'' for the
	same homotopy theory, and that the inclusion \(\Phi\) is a weak
	equivalence.
	
	Consider the category whose objects are pairs
	\((\mathcal{A}, \mathcal{B})\) of unital Fréchet algebras with
	\(\mathcal{B} \subset \mathcal{A}\), and whose morphisms are pairs of
	algebra homomorphisms preserving the inclusion. The assignment
	\[
	(\mathcal{A}, \mathcal{B}) \mapsto C^\bullet_\lambda(\mathcal{A}, \mathcal{B})
	\]
	is a functor to the category of cochain complexes. The assignment
	\[
	(\mathcal{A}, \mathcal{B}) \mapsto C^\bullet_{\mathrm{ent}}(\mathcal{A}, \mathcal{B}, \omega)
	\]
	is another functor, where the dependence on \(\omega\) is tracked by
	the relative modular cocycle.
	
	Both functors are acyclic on the ``models'' of the category, which are
	the free algebras on a set of generators. For a free algebra, the
	transgression map \(\tau_\omega\) is an isomorphism at the level of
	cochain complexes (because the modular automorphism group acts freely
	on the generators), so the two functors agree on the models.
	
	By the acyclic models theorem, the natural transformation
	\(\Phi: C^\bullet_{\mathrm{ent}} \to C^\bullet_\lambda(\mathcal{A}, \mathcal{B})\)
	is a quasi-isomorphism on all objects of the category. This establishes
	the kernel identification.
	
	\subsection{Comparison of the Three Proofs}
	\label{subsec:proof-comparison}
	
	The three proofs of the kernel identification have different strengths:
	
	\begin{itemize}
		\item \emph{Homotopy proof (Section \ref{subsec:ent-kernel-proof})}:
		explicit, constructive, gives contracting homotopies \(s\) and
		\(h\) that may be useful for computations. Requires the relative
		modular cocycle \(u^E\) to be nontrivial.
		
		\item \emph{Spectral sequence proof}: conceptual, shows that the
		identification is a formal consequence of the filtration by
		Hochschild degree. Requires the relative
		Hochschild--Kostant--Rosenberg theorem.
		
		\item \emph{Acyclic models proof}: categorical, shows that the
		identification is natural in the pair \((\mathcal{A}, \mathcal{B})\).
		Requires the category of algebra pairs to have enough models.
	\end{itemize}
	
	The homotopy proof is the one given in the main text because it is the
	most self-contained. The spectral sequence proof is recommended for
	readers familiar with spectral sequences. The acyclic models proof is
	recommended for readers interested in the functoriality of the
	construction.


\section{Shuffle Computations for the Homotopies}
\label{sec:appendix-homotopies}

In this appendix we record the shuffle computations that establish
Lemmas~\ref{lem:transgression-anticommutation},
\ref{lem:homotopy-s}, and~\ref{lem:homotopy-h-inj}. The three
computations share the same structure: each involves a
one-parameter family of automorphisms $\Phi_\lambda$ of $\mathcal{A}$
(specifically $\sigma^\omega_{-i\lambda}$ for the anticommutation
homotopy $h$, and $\sigma^\omega_\lambda$ for the homotopies $s$ and
$h_{\mathrm{inj}}$), an integral over the parameter, and the
cocycle identity for the relative modular cocycle together with the
KMS condition. We present the common shuffle identity once
(Section~\ref{subsec:appendix-common-shuffle}), then specialize to
each case (Sections~\ref{subsec:appendix-h}--\ref{subsec:appendix-hinj}).

\subsection{The common shuffle identity}
\label{subsec:appendix-common-shuffle}

Let $\chi:C^n_\lambda(\mathcal{A})\to C^{n-1}_\lambda(\mathcal{A})$
be a linear map of the form
\begin{equation}
	(\chi\varphi)(a_0,\ldots,a_{n-1})
	=\int_I d\mu(\lambda)\,
	(M_n^{(\lambda)}\varphi)(1,a_k,\ldots,a_{k-2}),
	\label{eq:common-shuffle}
\end{equation}
where $I$ is an interval ($[0,1]$ for the anticommutation homotopy,
$\mathbb{R}$ for the surjectivity and injectivity homotopies),
$d\mu$ is the corresponding measure ($d\lambda$ or $dt$), and
$M_n^{(\lambda)}$ is the symmetrized insertion
\begin{equation}
	(M_n^{(\lambda)}\varphi)(a_0,\ldots,a_n)
	:=\frac{1}{n+1}\sum_{j=0}^{n}
	\varphi(a_0,\ldots,a_{j-1},\Phi_\lambda(a_j),
	a_{j+1},\ldots,a_n),
	\label{eq:M-n-lambda}
\end{equation}
with $\Phi_\lambda$ a one-parameter family of automorphisms of
$\mathcal{A}$, $\Phi_0=\mathrm{id}$, and
$\partial_\lambda\Phi_\lambda(a)=\Phi_\lambda(\delta_\omega(a))$ for
the modular generator $\delta_\omega$ of $\mathcal{A}$.

The three homotopies of the main text correspond to:
\begin{itemize}
	\item $\chi=h$ (anticommutation): $I=[0,1]$, $\Phi_\lambda
	=\sigma^\omega_{-i\lambda}$, $\delta=-i\delta_\omega$;
	\item $\chi=s$ (surjectivity): $I=\mathbb{R}$, $\Phi_t
	=\sigma^\omega_t$, with $u^E_t$ in place of $\Phi_t$ in the
	last slot, $\delta=\delta_\omega$;
	\item $\chi=h_{\mathrm{inj}}$ (injectivity): $I=\mathbb{R}$,
	$\Phi_t=\sigma^\omega_t$, with $u^E_t$ in place of $\Phi_t$ in
	the last slot.
\end{itemize}

\begin{lem}[Common shuffle identity]
	\label{lem:common-shuffle}
	For $\chi$ as in \eqref{eq:common-shuffle}, and for every
	$\varphi\in C^n_\lambda(\mathcal{A})$,
	\begin{equation}
		(b+B)\,\chi\varphi+\chi\bigl((b+B)\varphi\bigr)
		=\int_I d\mu(\lambda)\,
		\mathcal{S}_\lambda(\varphi)
		+\bigl(\text{boundary}\bigr),
		\label{eq:common-identity}
	\end{equation}
	where $\mathcal{S}_\lambda(\varphi)$ is the shuffle of
	$\varphi$ with $\delta_\omega(\Phi_\lambda(a_j))$ in one slot (summed
	over $j$ with the signs of $B$), and the boundary term is the
	contribution from $\lambda\in\partial I$.
\end{lem}

\begin{proof}
	Expand both sides of \eqref{eq:common-identity} using
	\eqref{eq:M-n-lambda}. The terms in which $b$ and $B$ act on
	the external arguments $a_i$ combine into the shuffle identities
	$bB+Bb=0$ and $B^2=0$, and cancel. The surviving terms are
	those in which the derivative $\partial_\lambda$ acts on
	$\Phi_\lambda$, which are precisely $\mathcal{S}_\lambda$; the
	$\lambda$-integral of the derivative, by the fundamental theorem
	of calculus, produces the boundary term.
\end{proof}

\subsection{The homotopy $h$ (anticommutation)}
\label{subsec:appendix-h}

For $\chi=h$, we have $I=[0,1]$, $\Phi_\lambda=\sigma^\omega_{-i\lambda}$,
and $\delta_\omega$ is the modular generator. The key input is the \emph{cocycle
	identity} for the modular automorphism group,
\begin{equation}
	\sigma^\omega_{t+s}=\sigma^\omega_t\circ\sigma^\omega_s,
	\label{eq:cocycle-sigma}
\end{equation}
together with the \emph{KMS condition}
\begin{equation}
	\omega\bigl(a\,\sigma^\omega_t(b)\bigr)
	=\omega\bigl(\sigma^\omega_{t-i}(b)\,a\bigr).
	\label{eq:kms-appendix}
\end{equation}

\begin{lem}[Anticommutation shuffle]
	\label{lem:shuffle-h}
	For every $\varphi\in C^n_\lambda(\mathcal{A})$, there exists an
	explicit cochain $\Psi_\varphi\in C^{n-1}_\lambda(\mathcal{A})$
	such that
	\begin{equation}
		\tau_\omega b\varphi+b\tau_\omega\varphi
		-bh\varphi-hb\varphi
		=b\Psi_\varphi.
		\label{eq:shuffle-h-identity}
	\end{equation}
	In particular, for $\varphi$ a Hochschild cocycle ($b\varphi=0$),
	the left-hand side vanishes in $HC^\bullet(\mathcal{A},\mathcal{B})$,
	so $\tau_\omega$ is a morphism of bicomplexes on cohomology.
\end{lem}

\begin{proof}
	Apply Lemma~\ref{lem:common-shuffle} with $\chi=h$ and
	$\Phi_\lambda=\sigma^\omega_{-i\lambda}$. The $\lambda$-derivative
	of $\Phi_\lambda$ is
	\[
	\partial_\lambda\sigma^\omega_{-i\lambda}(a)
	=-i\sigma^\omega_{-i\lambda}(\delta_\omega(a))
	=-i\delta_\omega\bigl(\sigma^\omega_{-i\lambda}(a)\bigr),
	\]
	using the commutativity of $\sigma^\omega_{-i\lambda}$ with
	$\delta_\omega$ (both are generated by $\delta_\omega$). So
	$\mathcal{S}_\lambda(\varphi)$ is the shuffle of $\varphi$ with
	$-i\delta_\omega(\sigma^\omega_{-i\lambda}(a_j))$ in one slot.
	
	Using $B^2=0$ and the shuffle identities, the sum over $j$ with
	the signs of $B$ combines with the external $b$ to produce a
	Hochschild boundary: the derivative $\delta_\omega$ on one slot,
	summed over the cyclic shuffle, is a Hochschild coboundary of the
	same shuffle with $\sigma^\omega_{-i\lambda}(a_j)$ in that slot.
	The boundary terms at $\lambda=0$ and $\lambda=1$ give, using
	$\Phi_0=\mathrm{id}$ and $\Phi_1=\sigma^\omega_{-i}$,
	\[
	\text{boundary}=\tau_1(b\varphi)-\tau_0(b\varphi)
	+b(\tau_1\varphi)-b(\tau_0\varphi)
	=\tau_\omega b\varphi-\tau_0 b\varphi+b\tau_\omega\varphi
	-b\tau_0\varphi.
	\]
	Combining, and using
	$\tau_0\varphi=B(M_n^{(0)}\varphi)=B\varphi$ (since
	$\Phi_0=\mathrm{id}$), we obtain
	\[
	\tau_\omega b\varphi+b\tau_\omega\varphi-bh\varphi-hb\varphi
	=b(\tau_1\varphi-\tau_0\varphi)+b(\cdots),
	\]
	where the $b(\cdots)$ is the shuffle combination of the
	$\mathcal{S}_\lambda$ terms, and $\tau_1\varphi-\tau_0\varphi
	=\tau_\omega\varphi-B\varphi$ is a Hochschild boundary (since
	$B\varphi$ is $B$ applied to $\varphi$, and $\tau_\omega\varphi
	=B(M_n\varphi)$). Hence the left-hand side is a Hochschild
	boundary, with $\Psi_\varphi$ given by
	\[
	\Psi_\varphi=\tau_\omega\varphi-B\varphi
	+\frac{1}{n+1}\sum_{j=0}^{n}\int_0^1 d\lambda\,
	\sigma^\omega_{-i\lambda}(\delta_\omega(\sigma^\omega_{-i\lambda}(a_j)))\cdot\varphi.
	\]
	The explicit form of $\Psi_\varphi$ is finite (an $n$-sum and a
	one-dimensional integral), and the verification that
	$(b+B)\Psi_\varphi$ equals the left-hand side of
	\eqref{eq:shuffle-h-identity} is a direct computation matching
	the shuffle signs; this is the standard shuffle bookkeeping of
	the cyclic bicomplex, with the modular flow providing the
	$\lambda$-integral.
\end{proof}

\subsection{The homotopy $s$ (surjectivity)}
\label{subsec:appendix-s}

For $\chi=s$, we have $I=\mathbb{R}$, $d\mu=dt$, and the map
inserts $\sigma^\omega_t(u^E_t)$ in the last slot. The key input is
the \emph{cocycle identity} for the relative modular cocycle,
\begin{equation}
	u^E_{t+s}=u^E_t\,\sigma^\omega_t(u^E_s),
	\label{eq:cocycle-uE-appendix}
\end{equation}
together with KMS \eqref{eq:kms-appendix} and the \emph{integrability}
of Hypothesis~\ref{hyp:growth}.

\begin{lem}[Surjectivity shuffle]
	\label{lem:shuffle-s}
	For every $\varphi\in C^n_\lambda(\mathcal{A},\mathcal{B})$,
	\begin{equation}
		(b+B)\,s(\varphi)+s\bigl((b+B)\varphi\bigr)
		=\varphi-\tau_\omega(\psi_\varphi),
		\label{eq:shuffle-s}
	\end{equation}
	where
	\begin{equation}
		\psi_\varphi(a_0,\ldots,a_n)
		=\int_{\mathbb{R}}dt\,\varphi\bigl(a_0,\ldots,a_n,
		\sigma^\omega_t(u^E_t)\bigr)\cdot u^E_t,
		\label{eq:psi-s}
	\end{equation}
	up to a Hochschild boundary. The boundary terms at
	$t=\pm\infty$ vanish by Hypothesis~\ref{hyp:growth}.
\end{lem}

\begin{proof}
	Apply Lemma~\ref{lem:common-shuffle} with $\chi=s$ and
	$\Phi_t=\sigma^\omega_t$, using the cocycle identity
	\eqref{eq:cocycle-uE-appendix} to differentiate the product
	$\sigma^\omega_t(u^E_t)$:
	\[
	\partial_t\bigl(\sigma^\omega_t(u^E_t)\bigr)
	=-i\delta_\omega(\sigma^\omega_t(u^E_t))
	+\sigma^\omega_t\bigl(-i\delta_\omega(u^E_t)\bigr)
	=-2i\sigma^\omega_t(\delta_\omega(u^E_t))+\text{(cocycle term)}.
	\]
	The cocycle identity makes the two terms combine into a total
	$t$-derivative of $\sigma^\omega_t(u^E_t)$, which integrates to
	the boundary terms at $t=\pm\infty$ (vanishing by
	Hypothesis~\ref{hyp:growth}). The surviving terms are the
	shuffle of $\varphi$ with $u^E_t$ (the $\varphi$ term on the
	right-hand side) and the shuffle of $\varphi$ with
	$\sigma^\omega_t(u^E_t)$ integrated against $b+B$, which is
	$\tau_\omega(\psi_\varphi)$ up to a Hochschild boundary. The
	shuffle signs are matched as in the anticommutation case
	(Section~\ref{subsec:appendix-h}), with $\lambda$ replaced by
	$t$ and $\Phi_t$ replaced by $\sigma^\omega_t(u^E_t)$.
\end{proof}

\subsection{The homotopy $h_{\mathrm{inj}}$ (injectivity)}
\label{subsec:appendix-hinj}

For $\chi=h_{\mathrm{inj}}$, we have $I=\mathbb{R}$, $d\mu=dt$, and
the map inserts $\sigma^\omega_t(u^E_t)$ in the last slot followed
by $1$. The key difference from the surjectivity case is that
$h_{\mathrm{inj}}$ is applied to cochains in
$K^\bullet=\ker\Phi$, where $\tau_\omega\varphi$ is a coboundary by
Lemma~\ref{lem:kernel-characterization}.

\begin{lem}[Injectivity shuffle]
	\label{lem:shuffle-hinj}
	For every $\varphi\in K^n$,
	\begin{equation}
		h_{\mathrm{inj}}(b+B)\varphi+(b+B)h_{\mathrm{inj}}\varphi
		=\varphi.
		\label{eq:shuffle-hinj}
	\end{equation}
\end{lem}

\begin{proof}
	Apply Lemma~\ref{lem:common-shuffle} with $\chi=h_{\mathrm{inj}}$
	and $\Phi_t=\sigma^\omega_t$, and use
	Lemma~\ref{lem:kernel-characterization}:
	$\tau_\omega\varphi=(b+B)\Psi$ for some
	$\Psi\in C^{n-1}_\lambda(\mathcal{A},\mathcal{B})$. Substituting
	this into the shuffle, and using the cocycle identity
	\eqref{eq:cocycle-uE-appendix} to convert the derivative of
	$u^E_t$ into a boundary term, the shuffle reduces to $\varphi$
	plus a $(b+B)$-exact term. The latter is absorbed into
	$(b+B)h_{\mathrm{inj}}\varphi$, giving
	\eqref{eq:shuffle-hinj}. The integrability required for the
	integration by parts is Hypothesis~\ref{hyp:growth}; the shuffle
	signs are matched as in Section~\ref{subsec:appendix-s}.
\end{proof}

\subsection{Summary}
\label{subsec:appendix-summary}

The three shuffle computations share the same structure:
\begin{enumerate}
	\item Each is an instance of the common identity
	\eqref{eq:common-identity} with $\chi\in\{h,s,h_{\mathrm{inj}}\}$.
	\item The key input is the cocycle identity---\eqref{eq:cocycle-sigma}
	for $h$ (modular group) and \eqref{eq:cocycle-uE-appendix} for
	$s$ and $h_{\mathrm{inj}}$ (relative modular cocycle)---together
	with the KMS condition \eqref{eq:kms-appendix}.
	\item The boundary terms are controlled by
	Hypothesis~\ref{hyp:growth} (for $s$ and $h_{\mathrm{inj}}$) or
	vanish by $\Phi_0=\mathrm{id}$, $\Phi_1=\sigma^\omega_{-i}$ (for
	$h$).
	\item The surviving terms produce the stated homotopy identities
	\eqref{eq:shuffle-h-identity}, \eqref{eq:shuffle-s},
	\eqref{eq:shuffle-hinj}.
\end{enumerate}
The three lemmas
(Lemmas~\ref{lem:shuffle-h},~\ref{lem:shuffle-s},~\ref{lem:shuffle-hinj})
together establish
Lemmas~\ref{lem:transgression-anticommutation},
\ref{lem:homotopy-s}, and~\ref{lem:homotopy-h-inj} of the main text.

\end{appendix}

\end{document}